\documentclass[amsmath,amssymb,nofootinbib]{revtex4}
\usepackage{ae}
\usepackage[T1]{fontenc}
\usepackage[ansinew]{inputenc}
\usepackage{amsmath}
\usepackage{amssymb}
\usepackage{color}
\usepackage{graphicx}
\usepackage{subfigure}
\usepackage{longtable}
 \usepackage[usenames,dvipsnames]{pstricks}
\usepackage{epsfig}
\usepackage{pst-grad}
\usepackage{pst-plot}

\newcommand{\va}{\scriptscriptstyle}

\newcommand{\CS}{C_{\va \Sigma}}

\newcommand{\C}{\mathbb{C}}

\newcommand{\cg}{=_{{}_{\!\!\!\!\!\!\!c.g.}} }

\DeclareFontFamily{U}{rsfs}{}                                                                                          
\DeclareFontShape{U}{rsfs}{m}{n}{<5> rsfs5 <6><7> rsfs7          %
  <8><9><10><10.95><12><14.4><17.28><20.74><24.88> rsfs10}{}     %
\DeclareMathAlphabet{\mathfs}{U}{rsfs}{m}{n}                     %
\newcommand{\mfs}[1]{\mathfs {#1}}                               %

\newcommand{\sH}{{\mfs H}}

\newcommand{\sO}{{\mfs O}}

\newcommand{\Ha}{{\sH}_{aux}}

\newcommand{\be}{\nopagebreak[3]\begin{equation}}
\newcommand{\ee}{\end{equation}}
\newcommand{\ba}{\nopagebreak[3]\begin{eqnarray}}
\newcommand{\ea}{\end{eqnarray}}

\begin{document}
\scalebox{1}

\title{On the regularization of the constraints algebra of Quantum Gravity\\
 in $2+1$ dimensions with non-vanishing cosmological constant}
\author{Alejandro Perez\footnote{Unit\'e Mixte
de Recherche (UMR 6207) du CNRS et des Universit\'es Aix-Marseille
I, Aix-Marseille II, et du Sud Toulon-Var; laboratoire afili\'e
\`a la FRUMAM (FR 2291)}}
\email{perez@cpt.univ-mrs.fr}
\author{Daniele Pranzetti\footnote{
Universit\'e de Provence, Aix-Marseille I}}
\email{daniele.pranzetti@cpt.univ-mrs.fr}
\affiliation{Centre de Physique Th\'eorique,
Campus de Luminy, 13288 Marseille, France}

\begin{abstract}
\begin{center}
{\bf Abstract}
\end{center}
We use the mathematical framework of loop quantum gravity (LQG) to study the quantization of three dimensional (Riemannian) gravity with positive cosmological constant ($\Lambda>0$).  We show that the usual regularization techniques (successful in the $\Lambda=0$ case and widely applied in four dimensional LQG) lead to a deformation of the classical constraint algebra (or anomaly)  proportional to the local strength of the curvature squared. We argue that this is an unavoidable consequence of the non-local nature of generalized connections. \end{abstract}
\maketitle

\newpage \pagenumbering{arabic} %
\pagestyle{plain}

\section{Introduction}
Three dimensional quantum gravity can be defined from a number of different points of view. The first of these was the Ponzano-Regge \cite{Ponzano-Regge} model of quantum gravity on a triangulated 3-manifold which provides a quantization of  Regge calculus. The Ponzano-Regge model is a state sum model for 3-dimensional euclidean quantum gravity without cosmological constant using the Lie group $SU(2)$. Let $M$ be a triangulated compact 3-manifold, a state of the model is an assignment of an irreducible representation of $SU(2)$ to each edge of the triangulation. For each state there is a certain weight, a real number. The weight is given by the local formula:
\begin{equation}
W=\prod_{interior\,\,edges} (-1)^{2j}(2j+1)\prod_{interior\,\,triangles} (-1)^{j_1+j_2+j_3}\prod_{tetrahedra} \left\{
\begin{array}{ccc}
j_1 & j_2 & j_3 \\
j_4 & j_5 & j_6 \\
\end{array}
\right\}\,,
\end{equation}
where the weight for a tetrahedron is a $6j$-symbol and for each triangle the admissibility conditions are given by the requirement that
$j_1+j_2+j_3\in \mathbb{Z}$, while $j_1+j_2-j_3\geq 0$, $j_1+j_3-j_2\geq 0$, and $j_2+j_3-j_1\geq 0$.
The partition function, or state sum, is obtained by summing over all possible values of the spin on every edge in the interior of the manifold, subject to fixed values on the boundary:
\begin{equation}\label{eq:Partition Function}
Z=\sum_{j_1,j_2,..j_n}W\,.
\end{equation}
Since the set of irreducible representations of SU(2) is infinite, for some triangulations this gives a finite sum, whereas for some other triangulations this is a divergent infinite sum \footnote{One can show that this divergences are due to infinite contributions of pure gauge modes and eliminate them by appropriate gauge fixing. This procedure boils down to working with those triangulations where infinite sums are not present \cite{freidel-louapre}.}.

In the late eighties, early nineties mathematicians were looking for manifold invariants with the hope that this could help to classify them and this led Turaev and Viro \cite{Turaev-Viro} to the definition of a state sum which, under many aspects, was very similar to that of Ponzano and Regge. The two main differences were first that Turaev and Viro explicitly showed their model to be triangulation independent and secondly they replaced the Lie group $SU(2)$ with its quantum deformation $U_q SL(2)$. When the deformation parameter $q$ is a root of unity, then there are only a finite number of irreducible representations, which means that the edge lengths are not summed up to infinite values, and the partition function is always well-defined. A very important consequence of this is that the answer obtained is finite, and so the model appears to be a regularized version of the Ponzano-Regge model.

In particular, for $q=e^{\frac{i\pi}{r}}$, with integer $r\geq 3$, the weight $W_q$ for a given assignment of spins is given by a formula analogous to that for Ponzano-Regge in which every factor depends on q. Crucially, the spins are limited to the range $0 \leq j \leq (r - 2)/2$, and there is an extra admissibility condition
\begin{equation}
j_1+j_2+j_3\leq r-2\,.
\end{equation}
The $6j$-symbol is replaced by a quantum $6j$-symbol, and the factor for each edge by a quantum dimension. The Turaev-Viro model is defined by the partition function:
\begin{equation}
Z=N_q ^{-v}\sum_{j_1,j_2,..j_n=0}^{(r-2)/2}W_q\,,
\end{equation}
where $N_q$ is a constant depending on q, and $v$ is the number of internal vertices. Since the sum is finite, this is always well-defined. Moreover, the partition function $Z_q$ depends only on the boundary triangulation, the boundary data, and the topology of the manifold.

The natural question is then in what way the Turaev-Viro state sum is connected to Quantum Gravity?
The answer, first exhibited by Witten \cite{Witten}, and then rigorously proven by other authors (\cite{Walker}-\cite{Turaev}),
was that the Turaev-Viro state sum is equivalent to a Feynman path integral with the Chern-Simons action for
$SU(2)_k \otimes SU(2)_{-k}$, where $k$ is the level, and then the connection with gravity follows from the fact that the Chern-Simons action for this group product is related to the Einstein-Hilbert action for gravity with cosmological constant $\Lambda$ (\cite{Ooguri-Sasakura}, \cite{Williams}) if  $k^2 = 4\pi^2/\Lambda$.
Quantum groups also enter the quantization of Chern-Simons theory in the so-called combinatorial quantization approach (\cite{Combinatorial Quantization}), where a quantum deformation of the structure group is introduced as an intermediate regularization.

From the perspective of loop quantum qravity (LQG) only the case of vanishing cosmological constant is clearly understood. The quantization is in this case a direct implementation of Dirac's quantization program for gauge systems. The basic unconstrained phase space variables are represented as operators in an auxiliary Hilbert space (or kinematical Hilbert space $H_{kin}$ spanned by spin network states) where the constraints are represented by {\em regularized} quantum operators.  A nice feature of the regularization (which is both natural but also unavoidable in the context of LQG) is that it leads to regulated quantum constraints satisfying the appropriate quantum constraint algebra. There is no anomaly. This feature together with the background independent nature of the whole treatment allows for the removal of regulators and the definition of the physical Hilbert space and complete set of gauge invariant observables.

We give some additional details on the construction of the physical Hilbert space in the $\Lambda=0$ case, as they will allow us to relate the results of LQG in this simple case with what we have mentioned above concerning other approaches. The physical inner product and the physical Hilbert space $H_{phys}$ of $2+1$ gravity with $\Lambda = 0$ can be defined by introducing a regularization of the formal expression for the generalized projection operator into the kernel of curvature constraint (\cite{Rovelli}, \cite{Reisenberger}):
\begin{equation}\label{eq:Projector}
P=``\prod_{x\in \Sigma}\delta(\hat{F}(A(x))"=\int
D[N]exp\left(i\int_\Sigma Tr [N \hat{F}(A)]\right)\,.
\end{equation}
In \cite{Noui Perez} it has been shown how, introducing a regularization as an intermediate step for the quantization, this projector can be given a precise definition leading to a rigorous expression for the physical inner product of the theory which can be represented as a sum over spin foams whose amplitudes coincide with those of the Ponzano-Regge model. Moreover, the divergences mentioned above do not appear in the LQG treatment as the problematic triangulations (leading to infinities in the Ponzano-Regge state sum) simply do not contribute to physical transition amplitudes.

The previous discussion provides motivation to investigate the possible quantization, in the context of LQG, of the non vanishing cosmological constant case. If such program could be achieved one would expect to be able to understand
the Turaev-Viro amplitudes as the physical transition amplitudes or physical inner product between kinematical spin network states. Unfortunately, at the moment there is little evidence supporting this expectation.
Consequently, our work concentrates on the very basic starting point of the Dirac program: the study of the quantization of the constraints and their associated constraint algebra.  Our results reveals a puzzling feature that, we argue, is proper to the nature of the kind of regularizations admitted by the LQG mathematical framework, namely  the appearance of quantization anomalies.  As we will show, the nice interplay between symmetry and the regularizing nature of the LQG representation of basic kinematical observables,  present in the $\Lambda=0$ case, is lost in the case of non vanishing cosmological constant.
Thus, the generic appearance of anomalies in the quantum constraint algebra  appears  as a serious difficulty for the completion of the
LQG program for this model.

The paper is organized as follows:
In Section \ref{Ambiguities} we start with a brief discussion on the UV problem in LQG. The analysis of the constraint quantization and the associated algebra is presented in sections \ref{Classical Analysis}-\ref{Constraints Algebra}.

\section{UV divergences, their regularization, and ambiguities in LQG}\label{Ambiguities}

In standard background dependent QFT  in order to make sense of products of operator valued distributions one has to provide a regularization prescription. Removing the regulator is a subtle task involving the tuning of certain terms in the Lagrangian (counter terms) that ensure finite results when the regulator is removed. However, any of these regularization procedures is intrinsically ambiguous.
Ambiguities associated to the UV regularization allows to classify a theory as a renormalizable QFT if there are finitely many ambiguities which can be fixed by a finite number of renormalization conditions, i.e., one selects the suitable theory by appropriate tuning of the finite number of ambiguity parameters in order to match observations.

Removing UV divergences by a regularization procedure is intimately
related to the appearance of ambiguities in the quantum theory. This
problem is intrinsic to the formalism of QFT and is also present in
the context of LQG although in a disguised way \cite{Perez}\footnote{Even though this problem 
is a feature of quantum systems with local degrees of freedom it  
manifests itself also in loop quantum cosmology due to the peculiar nature of the
model whose definition uses the regulating-structures of the full theory\cite{lqc}.}.  In LQG
one aims at the canonical quantization of gravity in the connection
formulation. General relativity (Riemannian in 3d and both Riemannian
or Lorentzian in 4d) admits an unconstrained phase space formulation
in terms of an $SU(2)$ connection $A_a^i$ and its non Abelian
conjugate electric field $E^a_i$. At the unconstrained level, the
phase space is isomorphic to that of an $SU(2)$ Yang-Mills theory.
However, there is fundamental difference with Yang-Mills theory in
that general relativity contains diffeomorphisms as part of the gauge
group. This additional structure motivates the introduction of a
representation of the phase space basic variables as operators in an
auxiliary Hilbert space where diffeomorphisms are unitarily
represented. This important feature is crucial for the implementation
of diffeomorphism invariance (which in addition fixes uniquely---up to
unitarily equivalence---the representation of the basic unconstrained
(or kinematic) phase space variables as operators in a Hilbert space
\cite{hanno}). However, a prize to be paid with this choice of
quantization is the impossibility of representing the connection
$A_a^i$ as a quantum operator but only its holonomy (or parallel
transport) along one dimensional paths (the so called generalized
connection). In this way the basic quantum object is not the
configuration variable but a non local object constructed from it.  As
a consequence of this, local quantities entering the definition of the
constraints (such as the curvature tensor $F(A)$) do not admit a
direct quantum analog and must be replaced by expressions written in
terms of holonomies or Wilson loops along extended paths. This process
eliminates the potential UV divergences associated to non linearities
in the constraints---as it provides a natural point-split-like
regularization of local quantities---but generically introduces
infinite dimensional ambiguities. As explained in more detail below,
the reason for this is roughly the same that allows for infinitely
many possible ways of defining a lattice action for lattice Yang-Mills
theory when replacing the Yang-Mills connection by its parallel
transport in the lattice action.

The fact that these non linear quantities require regularization is
not surprising as it is a standard feature of quantum field theories
and the difficulty associated to the definition of products of
operator valued distributions at the same point. However, what is
special of LQG is that its kinematical structure obliges one to use a
regularization in terms of holonomies as these are the fundamental
quantum observables in the auxiliary Hilbert space (the generalized
connections).  Even the connection itself (the classical configuration
field) must be `approximated' by a {\em regulated} version in terms of
holonomies along infinitesimal paths. The fact that this process is
ambiguous can be illustrated by the following simple example.
Consider a one parameter family of contractible loops
$\alpha_{\epsilon}$ around the point $p\in \Sigma$ for $\epsilon>0$
such that the limit $\epsilon \to 0$ one has $\alpha_{\epsilon}\to
p$. If we denote by $W_{\epsilon}$ the holonomy of the connection $A$
around the loop $\alpha_{\epsilon}$, then the following is an infinite
dimensional family of regularizations of the local curvature
$\epsilon^{ab}F^i_{ab}(A)|_p$: \be f[\tau^i
W_{\epsilon}]=\frac{\sum\limits_{j=1/2}^{n/2} {c_j}
\frac{\chi_j(\tau^iW_{\epsilon})}{j(j+1) (2j+1)}}{\epsilon^2
\sum\limits_{k=1/2}^{n/2} c_k}, \label{genform}\ee for arbitrary
coefficients $c_j$ and some positive integer $n$, and where $\tau^i$
are the standard generators of the $su(2)$ Lie algebra, and
$\chi_{j}(g)$ is the character of $g\in SU(2)$ in the unitary
irreducible representation of spin $j$. The fact that for arbitrary
$c_j$'s and $n$ these are good regularizations of the curvature
component $\epsilon^{ab}F^i_{ab}(A)|_p$ at the point $p$ comes from
the fact that for smooth configurations of the connection one has
 \[f[\tau^i W_{\epsilon}]=\epsilon^{ab}F^i_{ab}(A)|_p+\sO(\epsilon^2),\]
independently of the freedom in choosing the parameters of the
family. Moreover, among the parameters of this family one should also
consider the functional arbitrariness entering in defining the above
homotopy of loops around $p$.  This is a particular example that
illustrates the nature of the ambiguities involved when replacing the
local connection by its parallel transport.

This (infinite dimensional) regularization ambiguity associated to the
need to replace connections by generalized connections are present in
the LQG quantization of three dimensional gravity for $\Lambda=0$. In
fact there is an infinite dimensional set of quantizations of the
constraints all of which are anomaly free at the regulated level. The
requirement of the quantum constraints to reproduce the appropriate
gauge symmetry algebra does not reduce the ambiguities in this
case. However, this by itself does not necessarily mean that the
quantum theory is ill defined: a detailed analysis \cite{Perez} shows
that when the regulator is removed in the construction of the
generalized projection operator $P$ (of equation (\ref{eq:Projector}))
the physical Hilbert space is uniquely determined: all physical
quantities are independent of the regularization ambiguities.

A second source of ambiguities, intimately related to the replacement of connections with (group valued) generalized connections, is that the conjugate momenta $E^a_i$ become non commutative \cite{zapata}. This introduces additional ordering ambiguities in cases where one is confronted with the quantization of non linear functionals of the electric field. These ambiguities are not present in the quantization of three dimensional gravity with vanishing cosmological constant as, in that case, all the constraints are at most linear in $E^a_i$. However, the ordering ambiguity will arise in the $\Lambda\not=0$ case where constraints are quadratic in $E^a_i$. We shall see that using the classical constraints algebra as a guiding principle the ordering ambiguities can indeed be reduced (see Section \ref{Constraints Algebra}).

Finally, there is the ambiguity associated to the details of the extended structures used for regulating the basic local quantities defining the constraints. For example, the shape of loops used in the definition of Wilson loops used to regulate the curvature strength, or the choice of co-dimension one surfaces used when the local $E^a_i$ field are replaced by flux operators admitting quantization in the LQG framework.  Although these ambiguities are not really relevant in the $\Lambda=0$ case, they have a strong effect on the constraint algebra in the $\Lambda\not=0$ case as well as in more general settings.

\section{Canonical three dimensional gravity with $\Lambda\neq 0$}\label{Classical Analysis}

We are interested in (Riemannian) three dimensional gravity with cosmological
constant in the first order formalism. The space-time $\mathcal{M}$ is
a three dimensional oriented smooth manifold and the action is
given by
\begin{equation}\label{eq:action}
S[e,\omega]=\int_{\mathcal{M}}{\rm tr}[e\wedge
F(\omega)+\frac{\Lambda}{3} e \wedge e\wedge e]
\end{equation}
\noindent where $e$ is a $\emph{su(2)}$ Lie algebra valued
$1$-form, $F(\omega)$ is the curvature of the three dimensional
connection $\omega$ and $Tr$ denotes a Killing form on
$\emph{su(2)}$. Assuming the space-time topology to be
$\mathcal{M}= \Sigma  \mathbb{R}$ where $\Sigma$ is a Riemann
surface of arbitrary genus, the phase space is parametrized by the
pull back to $\Sigma$ of $\omega$ and $e$. In local coordinates we
can express them in terms of the $2$-dimensional connection
$A^i_a$ and the triad field $E^b_j = \epsilon^{bc} e^k_c
\eta_{jk}$ where $a = 1, 2$ are space coordinate indices and $i, j
= 1, 2, 3$ are $\emph{su(2)}$ indices and $\epsilon^{ab}=-\epsilon^{ba}$ with $\epsilon^{12}=1$ (similarly $\epsilon_{ab}=-\epsilon_{ba}$ with $\epsilon_{12}=1$). The Poisson bracket among these variables is given  by
\begin{equation}\label{eq:symplectic}
\{A^i_a(x), E^b_j(y)\}=\delta^b_a \delta^i_j \delta^{(2)}(x,y)~.
\end{equation}
Due to the underlying $SU(2)$ and diffeomorphism gauge invariance
the phase space variables are not independent and satisfy the following set of first class constraints. The
first one is the analog of the familiar Gauss law of Yang-Mills theory, namely
\begin{equation}\label{eq:kinematic}
G_i\equiv D_a E^a_i=0~,
\end{equation}
\noindent where $D_a$ is the covariant derivative with
respect to the connection $A$. The constraint
(\ref{eq:kinematic}) is called the Gauss constraint. It encodes the condition that the connection be
torsion-less and it generates infinitesimal $SU(2)$ gauge
transformation. The second constraint reads
\begin{eqnarray}\label{eq:dynamic}
\nonumber C^i &=&\epsilon^{ab} (F_{ab}^i(A)+\Lambda \epsilon^{i}_{\ jk}e^j_a e^k_b)=0\\ &=& \epsilon^{ab} F_{ab}^i(A)+\Lambda \epsilon_{cd}\epsilon^{ijk}E^c_j E^d_k=0~,
\end{eqnarray} where in the first line we have written the constraint in terms of the triad field, while in the second line we have used the electric field.
This second set of first class constraints is associated to the diffeomorphism invariance of three dimensional gravity.
In order to exhibit the underlying (infinite dimensional) gauge symmetry Lie algebra it is convenient to smear  the constraints (\ref{eq:dynamic}) and (\ref{eq:kinematic}) with arbitrary test fields $\alpha$ and $N$, which we assume not depending on the phase space variables \footnote{In \cite{titi} one obtains a constraint algebra which mimics the constraint algebra of 4d gravity by using phase space dependent smearing functions. This choice considerably complicates the quantization of the (so defined) new constraints as their algebraic structure is no longer a Lie algebra.}, they read:
\begin{equation}\label{eq:kinematic_smeared}
G(\alpha)=\int_\Sigma \alpha^i G_i=\int_\Sigma \alpha^i D_a E^a_i=0
\end{equation}
and
\be\label{eq:dynamic_smeared} C(N)=\int_\Sigma N_i C^i=\int_\Sigma N_i(F^i(A)+\Lambda \epsilon^{ijk}E_j E_k)=0~.
\ee
The constraints algebra is then
\begin{eqnarray}\label{eq:constraints_algebra}
\nonumber \{C(N),C(M)\}&=& \Lambda \ G([N,M])\\ \nonumber \{G(\alpha),G(\beta)\}&=&G([\alpha,\beta])\\ \{C(N),G(\alpha)\}&=&C([N,\alpha]),
\end{eqnarray}
where $[a,b]^i=\epsilon^{i}_{\ jk} a^jb^k$ is the commutator of $su(2)$. For future use it will be convenient to split the constraint $C(N)$ as
\be
C(N)=F(N)+E( \Lambda N) \label{splitting1},
\ee
where
\be\label{splitting2}
F(N)=\int_\Sigma N_i(F^i(A)), \ \ \ \ \ \ \ \ \ \ \ \ \ E(\Lambda N)=\int_\Sigma \Lambda \epsilon^{ijk} N_i E_j E_k.
\ee
\subsection{Canonical Quantization}

The canonical quantization of the kinematics (i.e. the definition of the auxiliary Hilbert space where the constraints are to be quantized) is well understood.
Details can be found in \cite{bookt}.
Following Dirac's
quantization procedure one first finds a
representation of the basic variables in an auxiliary Hilbert
space $\Ha$. The key ingredient is the background independent construction of this auxiliary Hilbert space.
The main input is to replace functionals of the connection (taken as configuration variables) by functionals of holonomies along paths (the so-called generalized connections) $\gamma \subset
\Sigma$: these are the basic excitations in terms of which the Hilbert space is constructed. Given a connection $A$ and a path $\gamma$, one defines the
holonomy $h_{\gamma}[A]$ by
\be
\label{hol}h_{\gamma}[A]=P \exp\int_{\gamma} A \;.
\ee
The conjugate momentum (densitized triad) $E^a_i$ field is associated to its flux across co-dimension one surfaces. One
promotes these basic variables to operators acting on an auxiliary
Hilbert space where constraints are to be represented by quantum operators satisfying the
operator quantum analog of (\ref{eq:constraints_algebra}).

The auxiliary Hilbert space is defined by the Cauchy completion of
the space of cylindrical functionals ${Cyl}$, on the space of
(generalized) connections $\bar {\cal A}$\footnote{A generalized
connection is a map from the set of paths $\gamma \subset \Sigma$ to
$SU(2)$. It corresponds to an extension of the notion of holonomy
$h_{\gamma}[A]$ introduced above.}. The space $Cyl$ is defined as
follows: any element of $Cyl$, $\Psi_{\Gamma,f}[A]$, is a functional
of $A$ labeled by a finite graph $\Gamma \subset \Sigma$ and a
continuous function $f: SU(2)^{N_\ell({\Gamma})}\rightarrow \C$
where $N_\ell({\Gamma})$ is the number of links of the graph
$\Gamma$. Such a functional is defined as follows
\be \label{cyl}
\Psi_{\Gamma,f}[A]=f(h_{\gamma_1}[A],\cdots,h_{\gamma_{N_{\ell}(\Gamma)}}[A])
\ee
where $h_{\gamma_i}[A]$ is the holonomy along the link $\gamma_i$ of the graph $\Gamma$. { If one considers a new graph $\Gamma'$} such that $\Gamma \subset \Gamma^{\prime}$, then any cylindrical
function $\Psi_{\Gamma,f}[A]$ defined on $\Gamma$ can be promoted to a cylindrical function  $\Psi_{\Gamma^{\prime},f^{\prime}}[A]$ defined on $\Gamma^{\prime}$ in a natural way \cite{ash3}.
%
Given any two cylindrical functions
$\Psi_{\Gamma_1,f}[A]$ and $\Psi_{\Gamma_2,g}[A]$, their inner
product is defined by the Ashtekar-Lewandowski measure \ba
\label{innerk}\nonumber &&
<\Psi_{\Gamma_1,f},\Psi_{\Gamma_2,g}
> \equiv \mu_{AL}(\overline{\Psi_{\Gamma_1,f}[A]}\Psi_{\Gamma_2,g}[A])=\\ &&=\int \prod \limits_{i=1}^{N_{\ell_{\Gamma_{12}}}} dh_i \overline{f(h_{\gamma_1},\cdots,h_{\gamma_{N_\ell(\Gamma_{12})}})} g(h_{\gamma_1},\cdots,h_{\gamma_{N_\ell(\Gamma_{12})}})
\ea where $dh_i$ corresponds to the invariant $SU(2)$-Haar
measure, { $\Gamma_{12} \subset \Sigma$ is a graph containing both
$\Gamma_1$ and $\Gamma_2$, and we have used the same notation $f$
(resp. $g$) to denote the extension of the function $f$ (resp.
$g$) on the graph $\Gamma_{12}$}. The auxiliary Hilbert space
$\Ha$ is defined as the Cauchy completion of $Cyl$
under (\ref{innerk}).

The (generalized) connection is quantized  by promoting the holonomy (\ref{hol}) to an operator acting by multiplication in $\Ha$ as follows:
\be
\widehat{h_\gamma[A]} \Psi[A] \; = \; h_\gamma[A] \Psi[A]\;.
\label{ggcc}
\ee
The triad is associated with operators in $\Ha$ defining the flux of electric field across one dimensional lines. Namely, for a one dimensional path $\eta$ and a smearing field $\alpha:\Sigma\to su(2)$ we define
\be
E(\eta,\alpha)\equiv \int \alpha^i\hat{E}_i^a dx^b\epsilon_{ab} .\ee
The associated quantum operator in $\Ha$ can be defined from its action on holonomies. More precisely one has
\ba
\widehat{E(\eta,\alpha)}\triangleright h_{\gamma}[A]=\frac{i}{2}\ell_p \left\{\begin{array}{ccc} o_{\eta\gamma} \ \alpha h_{\gamma}[A]\ \ \ \ \mbox{(for $\eta$ target of $\gamma$)}\\ o_{\eta\gamma}\ h_{\gamma}[A]\alpha \ \ \ \ \mbox{(for $\eta$ source of $\gamma$)} \end{array}\right., \label{fluxx}
\ea
where the curve $\gamma$ is assumed to have one of its endpoints at $\eta$, $o_{\eta\gamma}=\pm1$ is the sign of the orientation of the pair of oriented curves in the order $(\eta,\gamma)$, and
where $\ell_p=\hbar G$ is the Planck length in three dimensions (the action vanishes if the curves are tangential to each other). In terms of
the triad operator we can construct geometric operators corresponding to the
area of regions in $\Sigma$ or the length of curves. The operators (\ref{ggcc}), and (\ref{fluxx}) are the basic extended variables in terms of which we shall regularize and quantize the constraints (\ref{eq:kinematic_smeared}) and (\ref{eq:dynamic_smeared}).
So far we have not specified the space of graphs that we are considering as this will also be part of the regularization procedure studied in the following section.

%

\section{Discretization prescription}\label{Discretization Prescription}

According to the canonical quantization program, we now need to introduce a quantization of the smeared constraints (\ref{eq:kinematic_smeared}), and (\ref{eq:dynamic_smeared})
preserving the gauge symmetry algebra at the quantum level. As we mentioned in previous sections, before quantizing it is necessary to translate the
classical variables entering the definition of the constraints in terms of holonomies of the connection and fluxes of the electric field.
In order to do that we need to define a discrete structure on top of which we can construct these extended variables. We do so by introducing  an
arbitrary finite cellular decomposition $\CS$ of $\Sigma$.  We denote $n$  the number of plaquettes
(2-cells) which from now on will be denoted by the index $p\in \CS$. We assume the plaquettes to be squares  with edges  (1-cells denoted $e\in \CS$) of  length $\varepsilon$ in a local coordinate system. It will also be necessary to use the dual complex $C_{\Sigma*}$ with its dual plaquettes $p*\in C_{\Sigma*}$ and edges $e*\in C_{\Sigma*}$ (see Fig. \ref{fig:Cellular_decomposition}). Both cellular decompositions inherit the orientation from the orientation of $\Sigma$.
The cellular decomposition defines the regulating structure. We now need to write the classical constraints in terms of extended variables in such a way that the naive continuum limit is satisfied.
Namely, it is necessary that (for smooth field configurations) the regulated classical constraints become the classical constraints in the limit $\epsilon\to 0$ (or equivalently $n\to \infty$).

Consequently, the phase space variables $E^a_i$ and $A_a^i$  are discretized as follows:
the local connection $A_a^i$ field is now replaced by the assignment of group elements $h(e)=P \exp(-\int_e A)\in SU(2)$ to the set of edges $e\in \CS$.
We discretize the triad field $E^a_i$ by assigning to each dual
1-cell $e*$ the $\emph{su(2)}$ element $E_i({e*})\equiv \int_{e*} \epsilon_{ab} E_i^b(x) dx^a$, i.e. the flux of electric field across the dual edge $e*$.
With this decomposition of $\Sigma$ we can write the regularized  versions of the constraints (\ref{eq:kinematic_smeared}) and (\ref{eq:dynamic_smeared}) as:
\begin{equation}\label{eq:kinematic_discretized}
G^{\va R}[\alpha]=\sum_{p*\in C_{\Sigma*}} {\rm tr}[\alpha^{p*} G^{p*}]=0
\end{equation}
and
\begin{equation}\label{eq:dynamic_discretized}
C^{\va R}[N]=\sum_{p\in\CS} {\rm tr}[N^p C^p],
\end{equation}
where $G^{p*}$ and $C^p$ are explicitly defined below.

Finally, we must define the set of states to be considered when studying the
regularized constraint algebra. The allowed states will be a subset $Cyl (\CS)\subset Cyl$
consisting of all cylindrical functions whose underlying graph is contained in the one-skeleton
of $\CS$. In other words, the allowed graphs must consist of collections of $1$-cells  $e\in\CS$.

\subsection{Regularized constraints algebra}\label{Constraints Algebra}

With the prescription introduced in the previous section we are now ready to compute the discrete version of the algebra (\ref{eq:constraints_algebra}).
Let us start with the sub-algebra of the Gauss constraint. Due to the simplicity of the action of $SU(2)$ gauge transformations on holonomies it is straightforward to obtain expressions for the regularized constraints that are anomaly free as far as it concerns the subalgebra of $SU(2)$ gauge transformations.
Thus the Gauss constraints  are more simply quantized by concentrating on their exponentiated versions. Namely, instead of writing the infinitesimal generator of $SU(2)$ gauge transformations as a self-adjoint operator acting on $\Ha$ it is simpler to directly construct an unitary operator generating finite $SU(2)$ gauge transformations. This follows from the fact that under a gauge transformation $g:\Sigma\to SU(2)$ the generalized connection transforms according to \be \label{gg}
h_{\gamma}[A] \longmapsto  g^{-1}_t h_{\gamma}[A]g_s
\ee
where $g_s=g(x_s),g_t=g(x_t)\in SU(2)$ are the value of the gauge transformation at the  {\em
source} and {\em target} points of $\gamma$ respectively.  Finite $SU(2)$ gauge transformations are then represented by the unitary operator $U_G(g)$ whose action on $\Psi_{\Gamma,f}\in Cyl$ is
\be \label{Ug}
U_G(g)\triangleright \Psi_{\Gamma,f}(\{h_{\gamma_i}\})\equiv \Psi_{\Gamma,f}(\{g^{-1}_{t^i}h_{\gamma_i}g_{s^i}\}).
\ee
The requirement that the quantization of the constraints satisfy the quantum counterpart of
 (\ref{eq:constraints_algebra}) translates into the following equations for the unitary generators
 \ba
U_G(g_1)U_G(g_2)=U_G(g_2g_1) .\label{qconst}
 \ea
The previous equations is satisfied by our definition and hence the canonical quantization
of the Gauss constraint presented here is {\em anomaly free}.
Given $\alpha:\Sigma \to su(2)$ one can compute the infinitesimal generator $\widehat G(\alpha)$  from the previous line as
\be
\widehat G(\alpha)=-i\hbar \frac{d}{dt} U_G(\exp(t\alpha))|_{t=0}. \label{infini}
\ee
Now the previous definition translates into the graphical action shown below which can be directly obtained by concentrating on a single graph node and formally expanding to first order in $\alpha$ equation (\ref{Ug}) where one has replaced $g=1+\alpha$. This allows us to define the action of ${\rm Tr}[\alpha^{p*} G^{p*}]$ appearing in (\ref{eq:kinematic_discretized}) where we have regularized the Gauss constraint with a sum over dual plaquettes. In fact, since each dual plaquette correspond to a single node of the cellular decomposition, we can define the action of ${\rm Tr}[\alpha^{p*} G^{p*}]$ on the node (to which the dual plaquette $p*$ corresponds) which is target of four holonomies using the following graphical notation:
\ba&&
{\rm Tr}[\alpha^{p*} G^{p*}]\rhd h_1\otimes h_2\otimes h_3\otimes h_4=\\
&& \nonumber  \cg \begin{array}{c}  \includegraphics[width=3.4cm,angle=360]{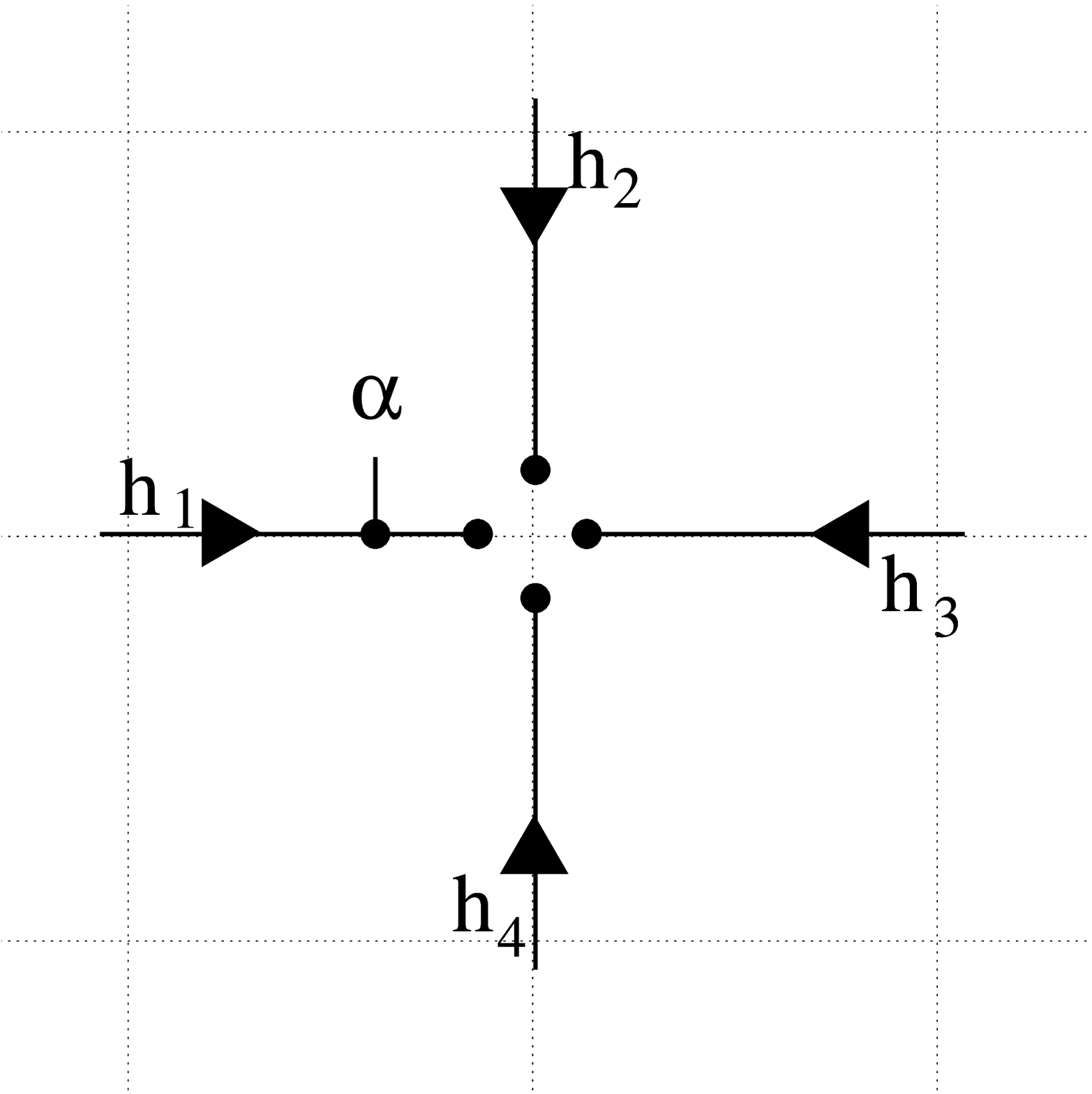}\end{array} + \begin{array}{c}\includegraphics[width=3.4cm,angle=360]{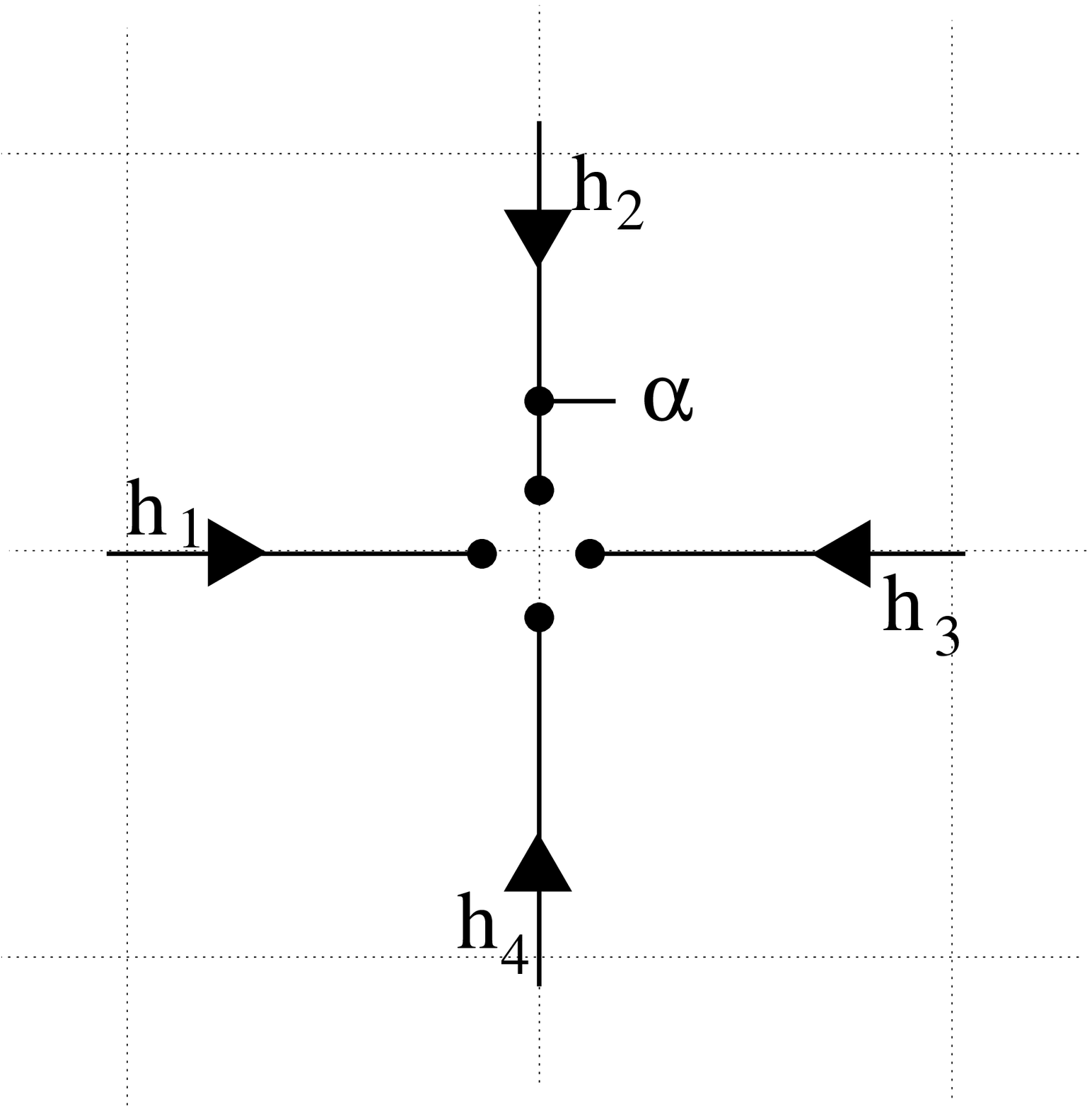}\end{array}+\begin{array}{c}  \includegraphics[width=3.4cm,angle=360]{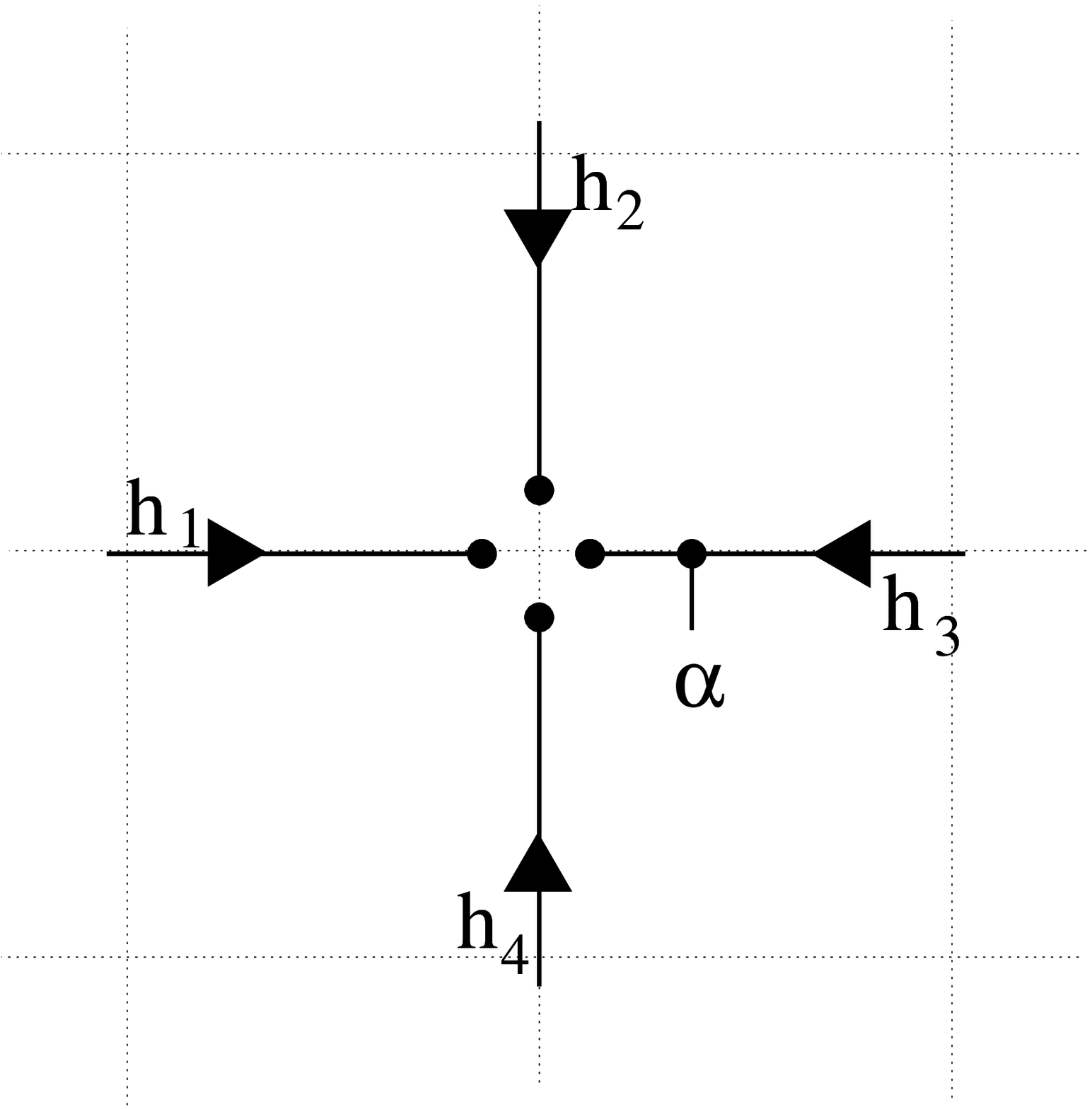}\end{array} + \begin{array}{c}\includegraphics[width=3.4cm,angle=360]{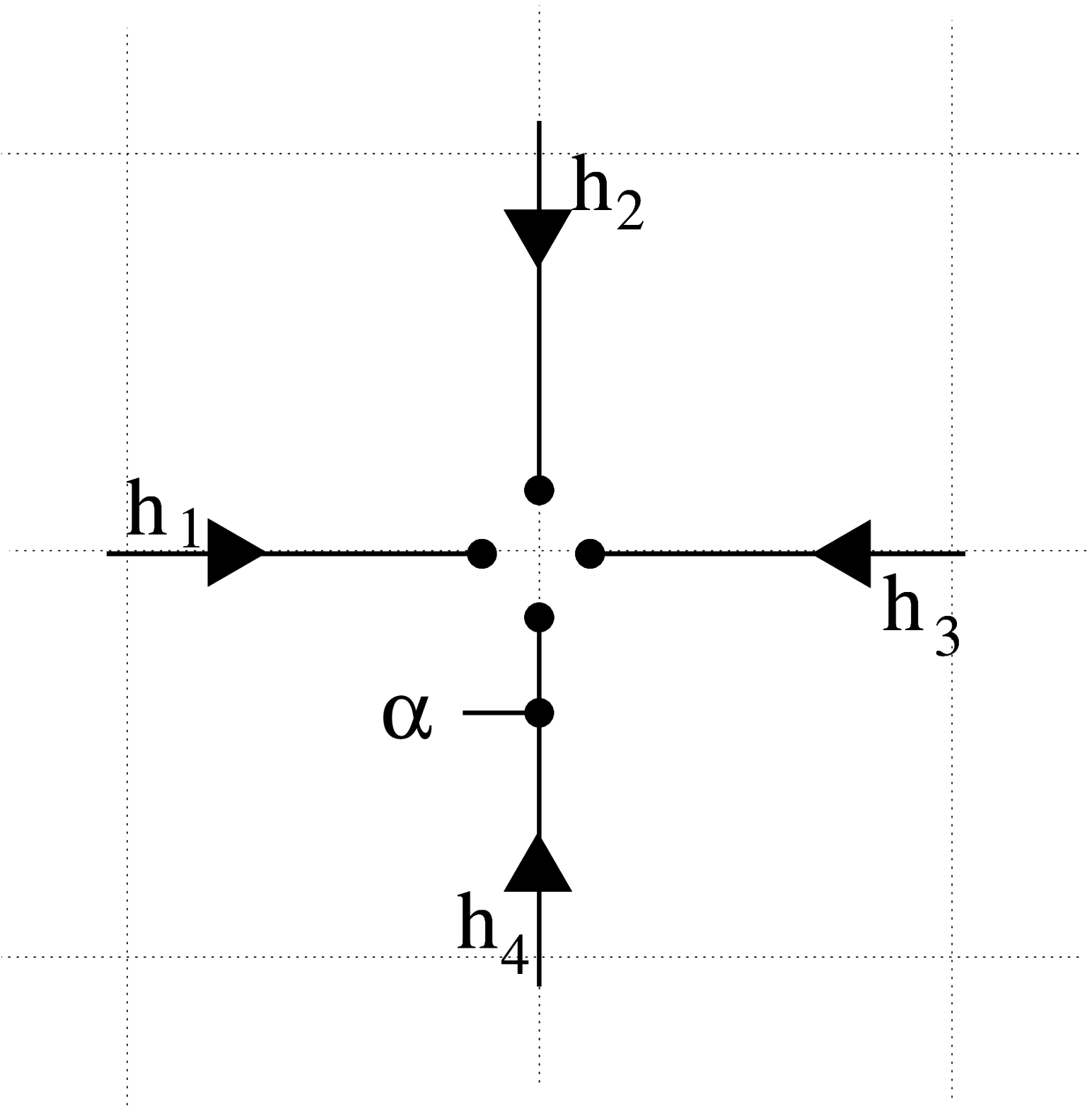}\end{array}\ea
where we have omitted the index ${p*}$ for the smearing field in the figure. Notice that from (\ref{fluxx}) and the last line, the action of the Gauss constraint on the node can be interpreted as the flux operator $E(C,\alpha)$ across an infinitesimal circle $C$ (a zero area circle) centered at the node which is reminiscent of the well known geometric interpretation of the Gauss law in the Abelian case such as electromagnetism.

The other cases corresponding to all possible orientations of the four holonomies can be obtained from
(\ref{fluxx}) or from (\ref{infini}) in a similar fashion. The action of the Gauss constraint on arbitrary elements of $Cyl(\CS)$ can be  obtained from the previous equations by the standard rules of differentiation of functions of finitely many copies of $SU(2)$.

From the action of the Gauss constraint in a given dual plaquette above it is immediate to see how the Gauss constraint sub-algebra is preserved also at the quantum level without anomalies, as we expect from equations (\ref{qconst}) and (\ref{infini}). Infact, due to the local action of the Gauss constraint it is sufficient to concentrate on the action of the commutator $[G^{\va R}(\alpha),G^{\va R}(\beta)]$ on a single node. At the regularized level this means that  only the dual plaquette around the given node is really relevant and so, among all the terms in the commutator due to the sum over dual plaquettes in the definition of $G^{\va R}$, only one gives a non vanishing contribution, namely:
\begin{eqnarray}
&&[G^{\va R}(\alpha),G^{\va R}(\beta)]\rhd\Psi
=[Tr[\alpha^{p*} G^{p*}],Tr[\beta^{p*} G^{p*}]]\rhd
h_1\otimes h_2\otimes h_3\otimes h_4=\nonumber\\
&=&\bigg(([\alpha^{p*}, \beta^{p*}]\,h_1)\otimes h_2\otimes h_3\otimes h_4 +h_1\otimes([\alpha^{p*}, \beta^{p*}]\,h_2)\otimes h_3\otimes h_4+\nonumber\\
&+&h_1\otimes h_2\otimes([\alpha^{p*}, \beta^{p*}]\,h_3)\otimes h_4+h_1\otimes h_2\otimes h_3\otimes([\alpha^{p*}, \beta^{p*}]\,h_4)\bigg)=\nonumber\\
&=& G^{\va R}([\alpha, \beta])\rhd\Psi~.
\end{eqnarray}
Hence, the quantum Gauss constraints reproduce the correct commutator algebra and are thus anomaly free as far as the $SU(2)$ sub algebra is concerned.
This property is related to the well known, and extremely useful fact in the context of lattice gauge theories,  that discretization does not break the
Yang-Mills gauge symmetry.

\begin{figure} \centerline{\hspace{0.5cm}\(
\begin{array}{ccc}
\includegraphics[height=4cm]{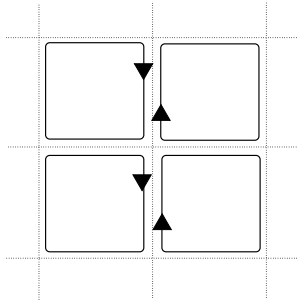}
\end{array}
\)}
\caption{Orientation of plaquette holonomies chosen for the regularization of $F[N]$.}
\label{curvature}
\end{figure}

We shift now the attention to the definition of the regulated constraints associated with (\ref{eq:dynamic_discretized}). According to (\ref{splitting1}) and (\ref{splitting2}), we can write the regulated constraint as
\be C^{\va R}[N]=F^{\va R}[N]+E^{\va R}[\Lambda N]\ee
which allows us in what follows to consider the regularization of $F[N]$ and $E[\Lambda N]$ separately.
The main observation that motivates the regularization of $F[N]$ is that the integral defining this first term can be approximated by a Riemann sum over plaquette contributions according to  $F[N]=\int_\Sigma {\rm tr}[NF(A)]= \lim_{\varepsilon\rightarrow 0}\sum_p \varepsilon^2{\rm tr}[N^p F^p]$. The Riemann sum is then the regulated quantity to be promoted to a quantum operator. We only need now to approximate  the curvature tensor by an expression constructed in terms of holonomies of the connection along suitable paths in such a way that the latter expression converges to the local curvature of the connection $A$ (for smooth configurations of $A$) in the limit $\epsilon \to 0$. To this end we define the holonomies $g_e\in SU(2)$ associated to  each oriented edge $e\in \CS$. In terms of this definition the holonomy around a single plaquette $p\in \CS$ becomes
$W^p=g_{e_1^p}\cdots g_{e_4^p}$ where here $e_i^p$ for $i=1,\cdots, 4$ are the corresponding edges bounding the plaquette of interest. Finally using that (for smooth configurations) $W^p[A]= \mathbf{1} +\varepsilon^2 F^p[A]+\mathcal{O}(\varepsilon^2)$ a natural candidate for regulated $F[N]$ is
\be F^{\va R}[N]=\sum_p F[N^p]=\sum_p {\rm tr}[N^p (W^p)].\label{curvy}\ee
Notice that the previous regularization corresponds to the choice $n=1$ at each plaquette in the general formula given in equation (\ref{genform}). At the end of this paper we will discuss the
generalization of our analysis to arbitrary regularizations. We will argue that the main result of this work is indeed generic and thus independent of the choice made here for the sake of simplicity.
\begin{figure} \centerline{\hspace{0.5cm}\(
\begin{array}{ccc}
\includegraphics[height=4cm]{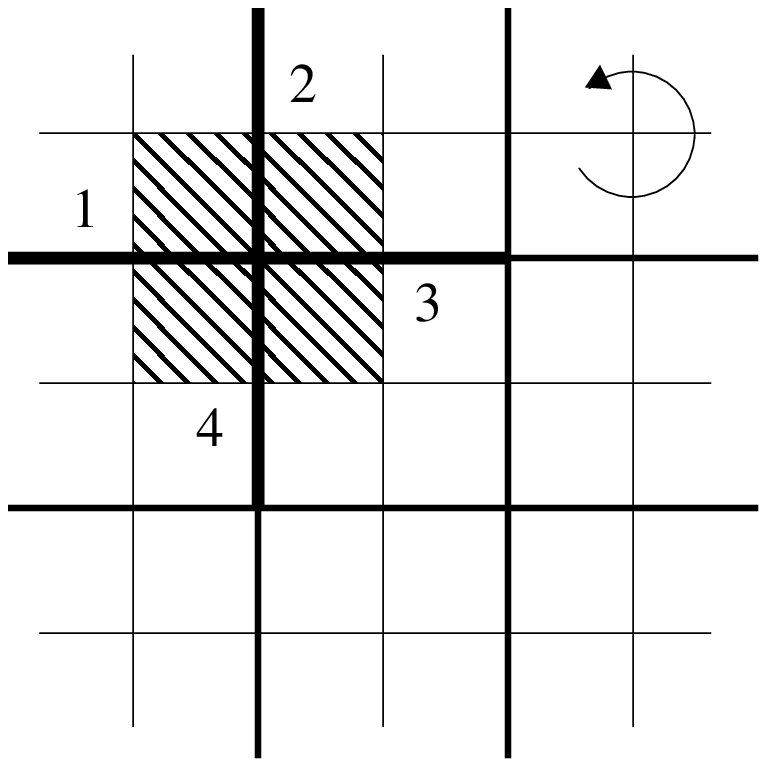}
\end{array}\ \ \ \ \ \ \ \ \
\begin{array}{ccc}
\includegraphics[height=1cm]{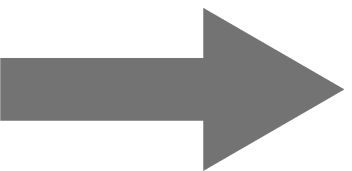}
\end{array} \ \ \ \ \ \ \ \ \
\begin{array}{ccc}
\includegraphics[height=4cm]{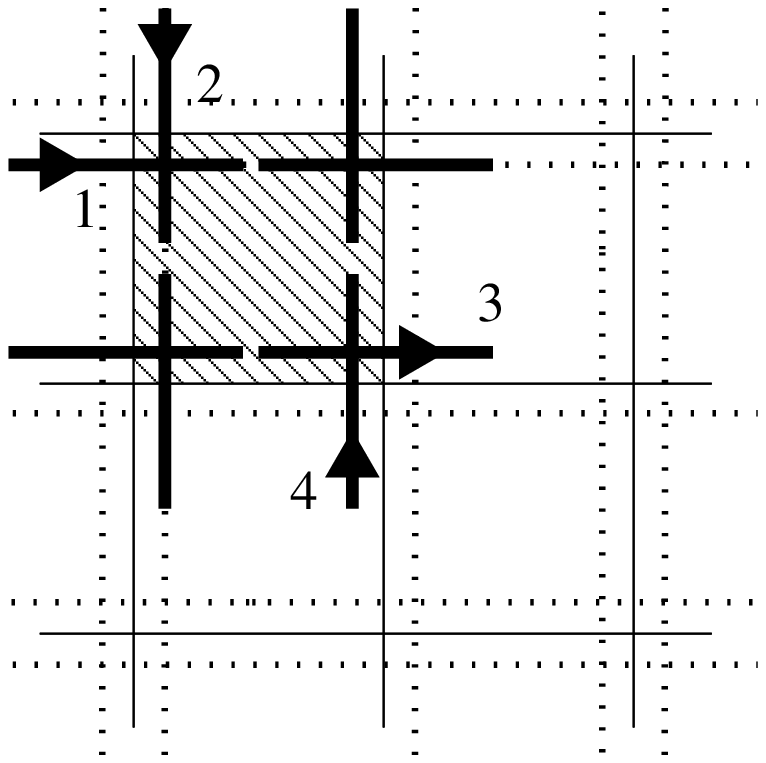}
\end{array}
\)} \caption{On the left: portion of the cellular decomposition $\CS$ (thin lines) and its dual $\CS^{\star}$ (thick lines). On the right: the edges of $\CS^{\star}$ are shifted toward the corresponding nodes. The flux operators necessary for the definition of the regularization of $E[\Lambda N]$ are defined in terms of the latter shifted dual edges.}
\label{fig:Cellular_decomposition}
\end{figure}

\begin{figure} \centerline{\hspace{0.5cm}\(
\begin{array}{ccc}
\includegraphics[height=3.6cm]{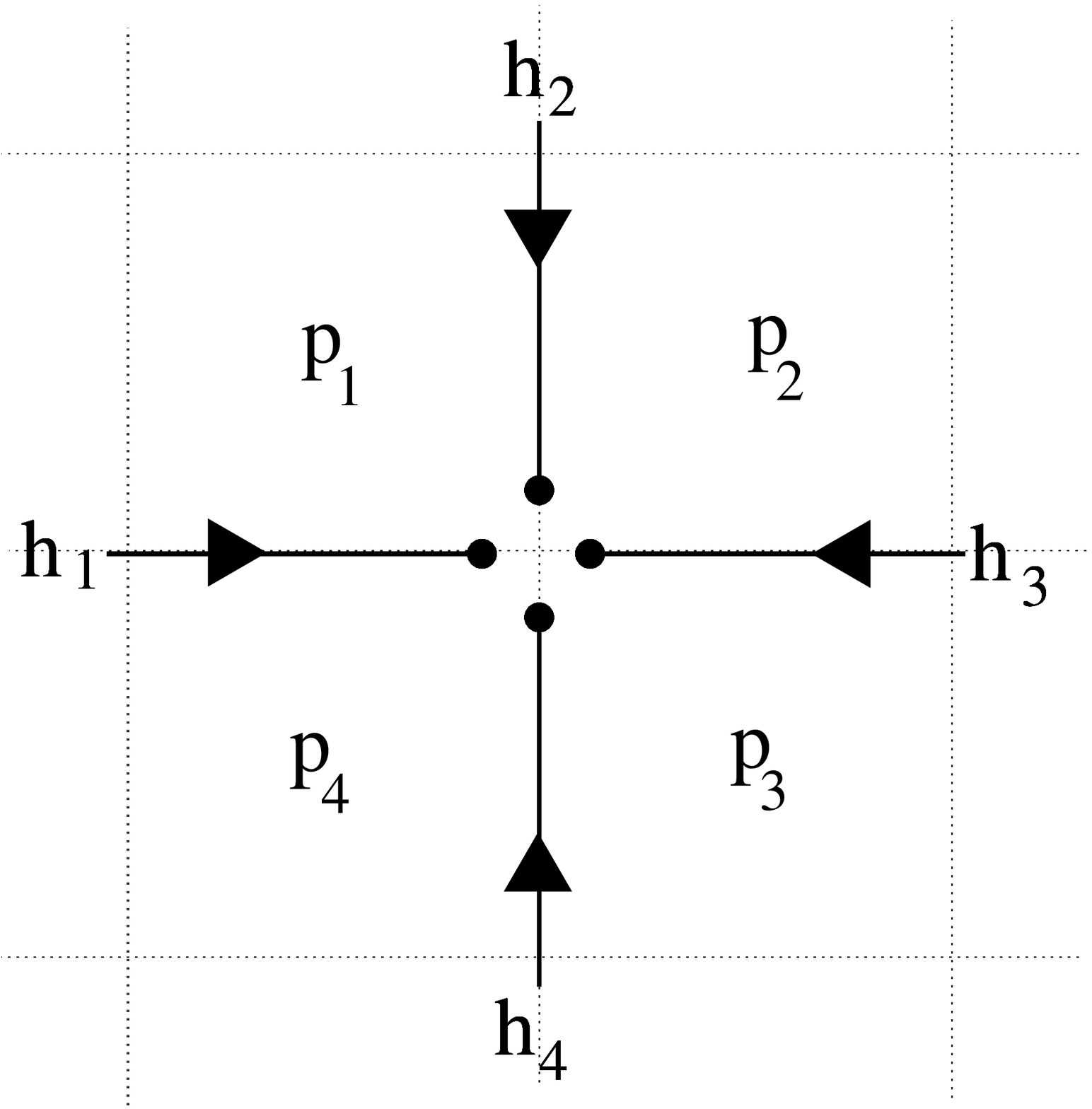}
\end{array}\ \ \ \ \ \ \ \ \
\begin{array}{ccc}
\includegraphics[height=3.6cm]{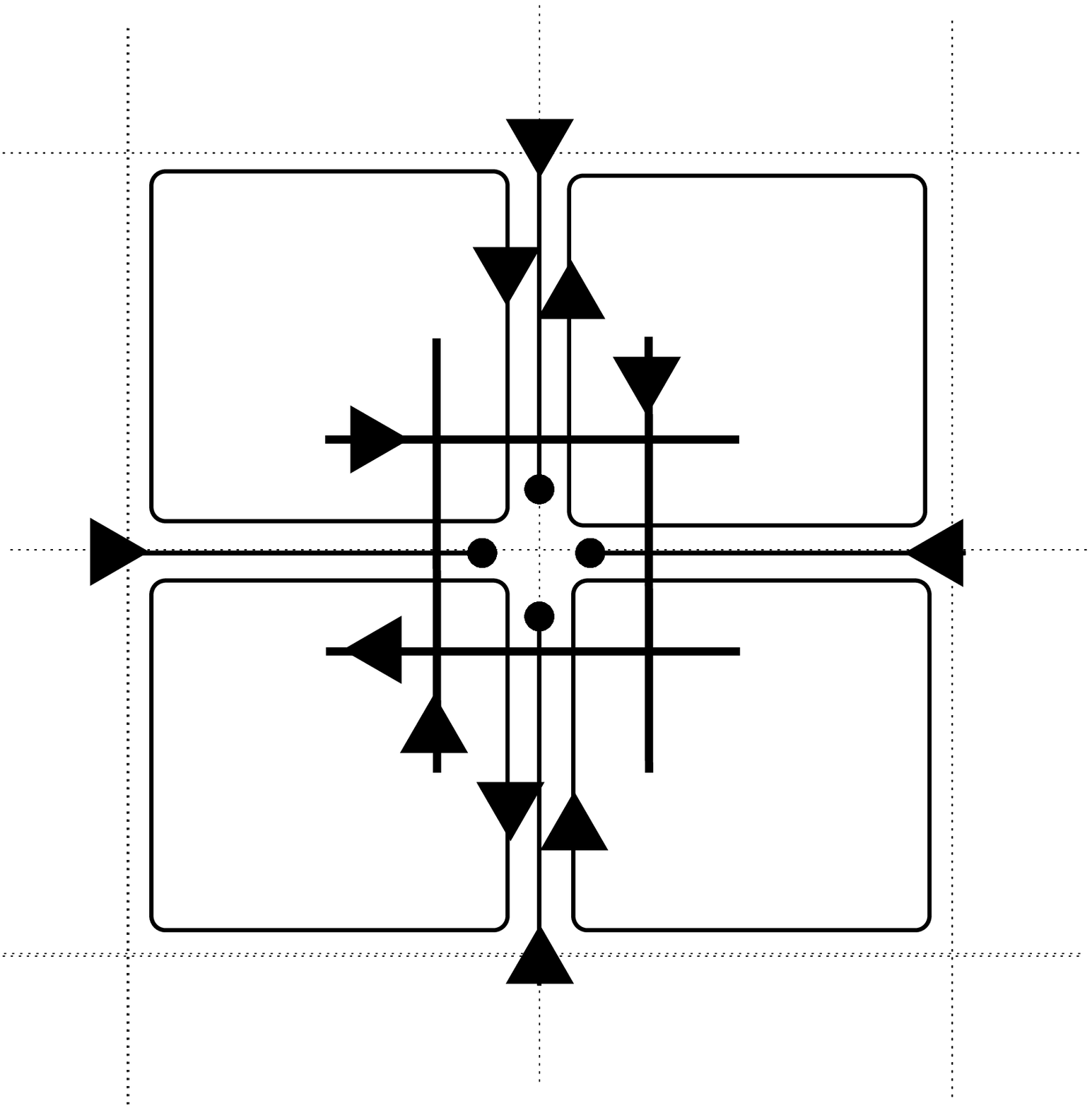}
\end{array} \)}\caption{On the left: a  generic non $SU(2)$ invariant state $\Psi$ in our cellular decomposition. On the right: explicit illustration of the orientation of the paths used in the regularization of the constraints.}
\label{fig:State}
\end{figure}

\begin{figure} \centerline{\hspace{0.5cm}\(
\begin{array}{ccc}
\includegraphics[height=1.2cm]{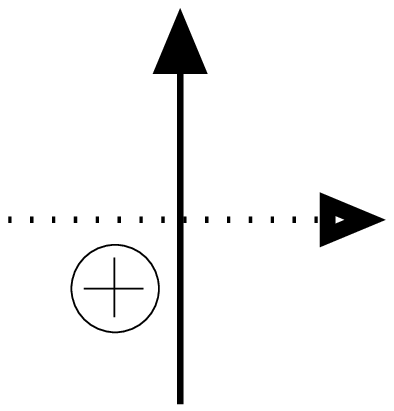}\\
\; \, \includegraphics[height=1.2cm]{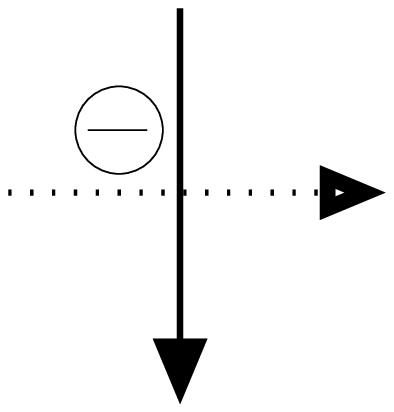}
\end{array}\ \ \ \ \ \ \ \ \
\begin{array}{ccc}
\includegraphics[height=1.2cm]{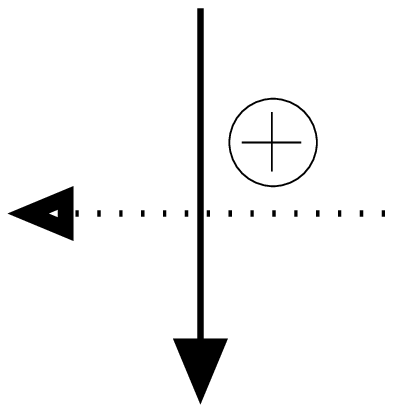}\\
\includegraphics[height=1.2cm]{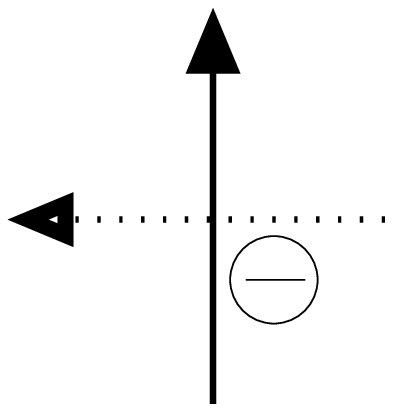}
\end{array}\ \ \ \ \ \ \ \ \
\begin{array}{ccc}
\includegraphics[height=1.2cm]{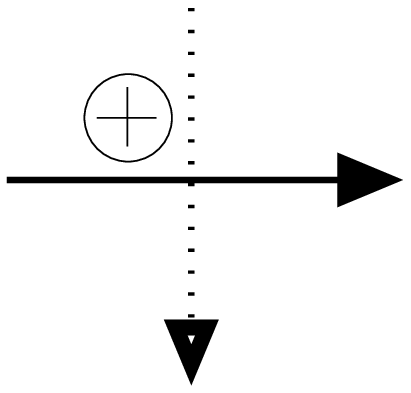}\\
\includegraphics[height=1.2cm]{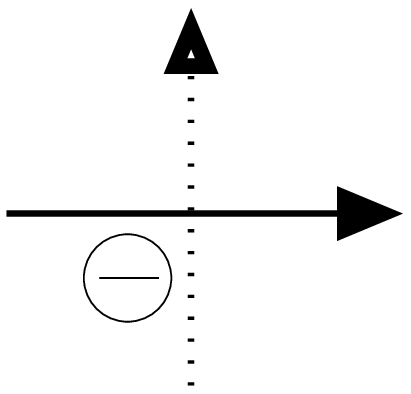}
\end{array}\ \ \ \ \ \ \ \ \
\begin{array}{ccc}
\includegraphics[height=1.2cm]{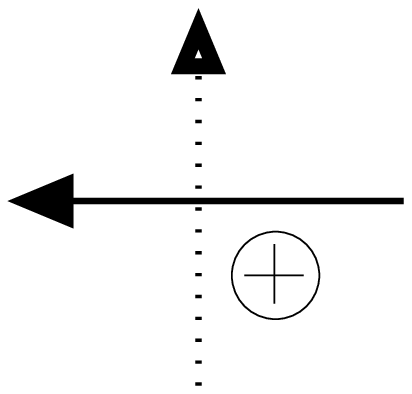}\\
\; \, \includegraphics[height=1.2cm]{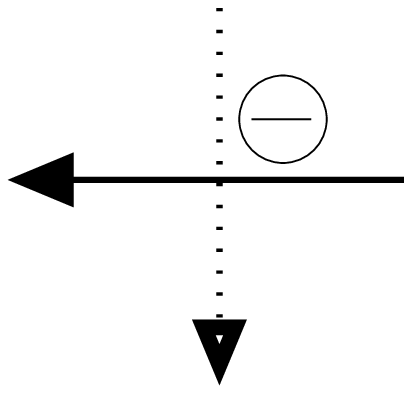}
\end{array}  \)}\caption{All possible values of the orientations $o_{\eta\gamma}$ of the crossing appearing in equation (\ref{fluxx}).  The edges $\eta$ regulating the flux operators are the dotted lines while the holonomies are along the continue lines $\gamma$.}
\label{fig:Signs}
\end{figure}

The regularization of $E[\Lambda N]$ is more subtle as one needs to replace the classical smooth field $E^a_i$ by extended flux operators which,
according to (\ref{fluxx}), act as `grasping' operators.  An important observation, that follows from the action of the regulated Gauss constraint,
is that the grasping operators must act at the endpoints of the edge holonomies in order to avoid inconsistency. Therefore, instead of smearing the $E$-field along the edges of the usual
dual cellular decomposition $\eta\in \CS^{\star}$ we need to work with the flux of $E$-field associated with the shifted edges depicted on the right of Figure \ref{fig:Cellular_decomposition}.
By abuse of notation we shall keep denoting the shifted edges $\eta\in \CS^{\star}$.
We write the regulated quantity corresponding to $E[\Lambda N]$ as
\be
E^{\va R}[\Lambda N]= \sum_{p} E[\Lambda N_p],
\ee
where (concentrating on the shadowed plaquette in Figure  \ref{fig:Cellular_decomposition})
\begin{eqnarray}
E[\Lambda N_p] &=&\Lambda \epsilon_{ijk} N^i_p \ ( E(\eta_1,\tau^j)E(\eta_2,\tau^k))+\Lambda \epsilon_{ijk} N^i_p\  ( E(\eta_2,\tau^j)E(\eta_3,\tau^k))\nonumber\\
&+&\Lambda \epsilon_{ijk} N^i_p \ ( E(\eta_3,\tau^j)E(\eta_4,\tau^k))+\Lambda \epsilon_{ijk} N^i_p\  ( E(\eta_4,\tau^j)E(\eta_1,\tau^k))\,,
\end{eqnarray}
where $\eta_{i}\in \CS^{\star}$ are the four shifted edges shown in the figure that are dual to the shadowed plaquette
$p\in \CS$, and the operators $E(\eta,\alpha)$ are defined in (\ref{fluxx}).
Let us start with the quantum version of
\begin{equation}\label{eq:dinamic}
\{C[N],C[M]\}= \Lambda G(\{N,M\})~.
\end{equation}
In order to compute the discrete analog of (\ref{eq:dinamic}) it is sufficient to derive the action of the r.h.s and l.h.s of this commutation relation around a single vertex, more precisely around a generic non $SU(2)$-invariant state at a vertex of the cellular decomposition, since both sides of the commutator (\ref{eq:dinamic}) vanish for $N^p$, $M^{p'}$ belonging to two plaquettes $p$ and $p'$ that do not share a common vertex. Therefore the quantum version of eq. (\ref{eq:dinamic}) has to be computed for a state of the form shown in Fig. \ref{fig:State} and the sums over plaquettes inside the expressions of $C^{\va R}[N]$ and $C^{\va R}[M]$ give not vanishing contributions only for a finite number of plaquettes, precisely the ones around the given vertex.\\
\\
Let us call this non $SU(2)$ invariant state $\Psi$ and write the discrete action of the l.h.s of (\ref{eq:dinamic}) on $\Psi$:
\begin{eqnarray}\label{eq:dinamic_discrete_left}
&& [C^{\va R}[N],C^{\va R}[M]]\!\!\rhd\!\! \Psi =\nonumber \\ &&= [[C(N^1), C(M^1)]+[C(N^1), C(M^2)]+[C(N^1), C(M^3)]+[C(N^1), C(M^4)]+\nonumber\\
&&+[C(N^2), C(M^2)]+[C(N^2), C(M^1)]+[C(N^2), C(M^3)]+[C(N^2), C(M^4)]+\nonumber\\
&&+[C(N^3), C(M^3)]+[C(N^3), C(M^1)]+[C(N^3), C(M^2)]+[C(N^3), C(M^4)]+\nonumber\\
&&+[C(N^4), C(M^4)]+[C(N^4), C(M^1)]+[C(N^4), C(M^2)]+[C(N^4), C(M^3)]]\!\!\rhd\!\! \Psi.
\end{eqnarray}
%
%
We have now to choose also a convention of signs for the action of each $E$ field on holonomies, we pick the one shown in Fig. \ref{fig:Signs}, where the dashed arrows represent the $E$ field and the black ones holonomies.
With these conventions we are now ready to compute the action (\ref{eq:dinamic_discrete_left}). Let us start with the terms involving the commutator of regulated quantities at the same plaquette. For the plaquette $1$ we have:\\
\begin{eqnarray*}
&& [C(N^1), C(M^1)]\rhd \Psi=\left([\left[F[N^1],E[\Lambda M^1]\right]+\left[E[\Lambda N^1],F[M^1] \right]\right)\rhd \Psi=\nonumber\\
&&=\frac{1}{2} \begin{array}{c}  \includegraphics[width=3.4cm,angle=360]{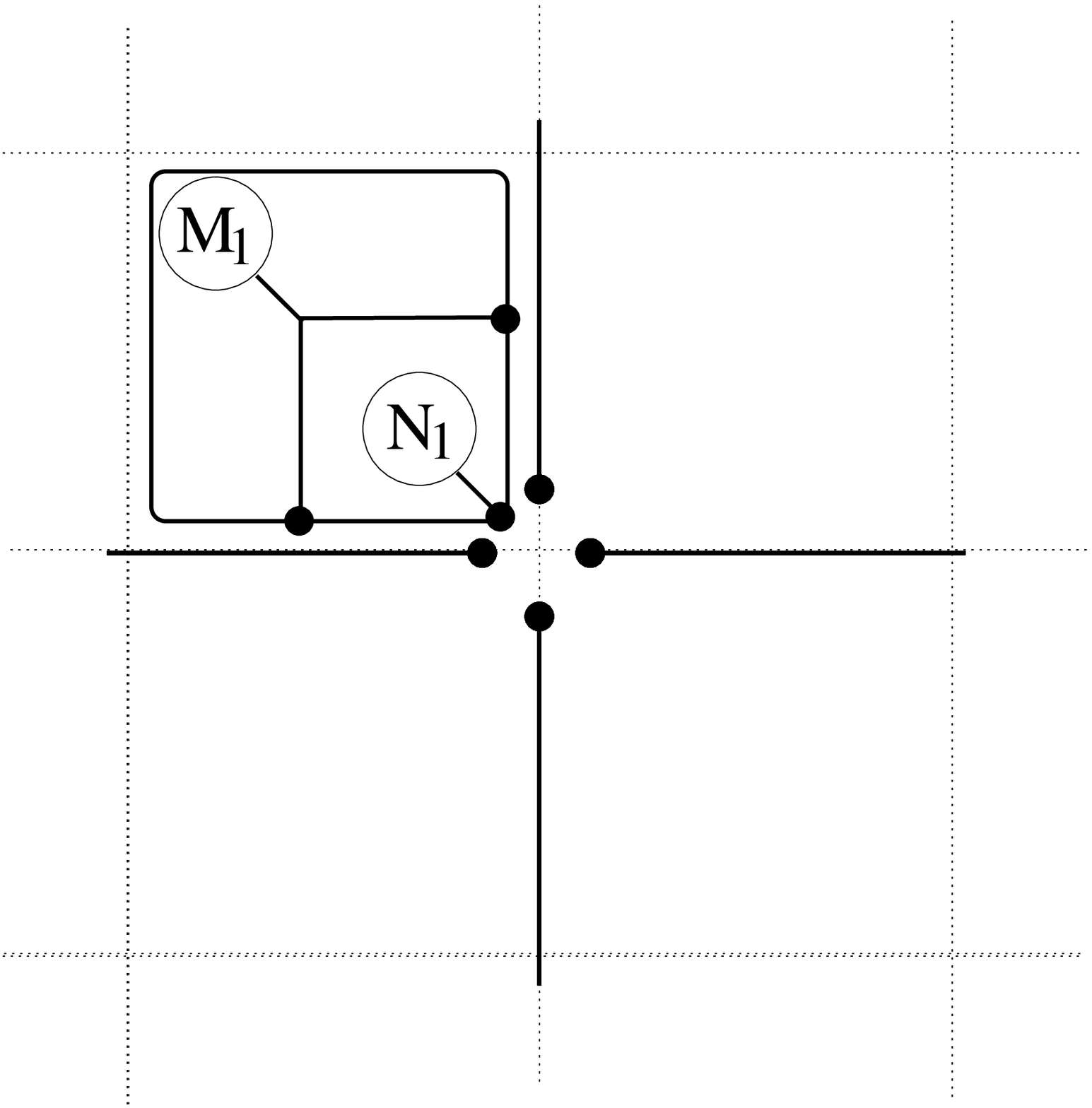}\end{array}+\frac{1}{4}\begin{array}{c}\includegraphics[width=3.4cm,angle=360]{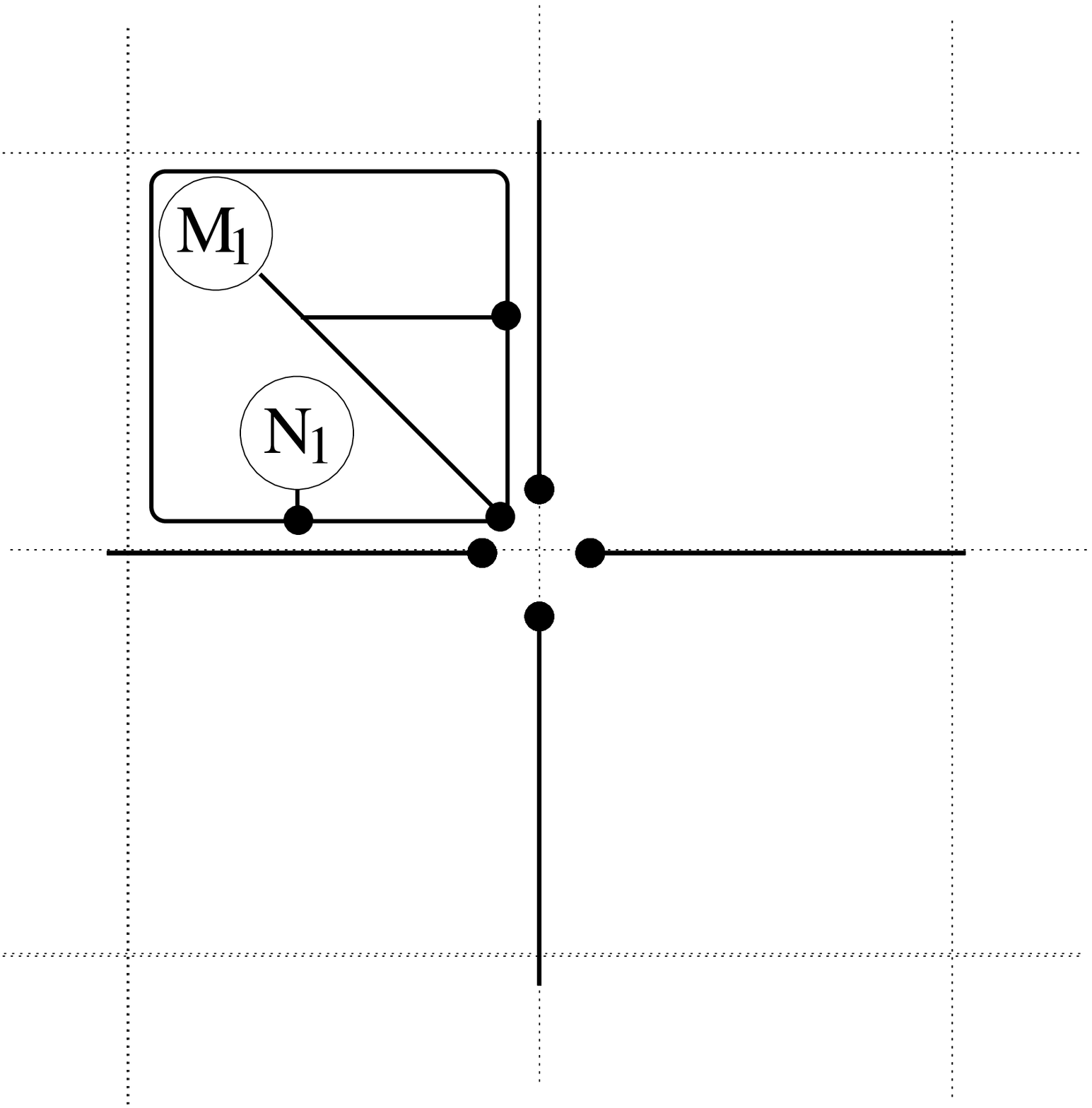}\end{array}+\frac{1}{4}\begin{array}{c}\includegraphics[width=3.4cm,angle=360]{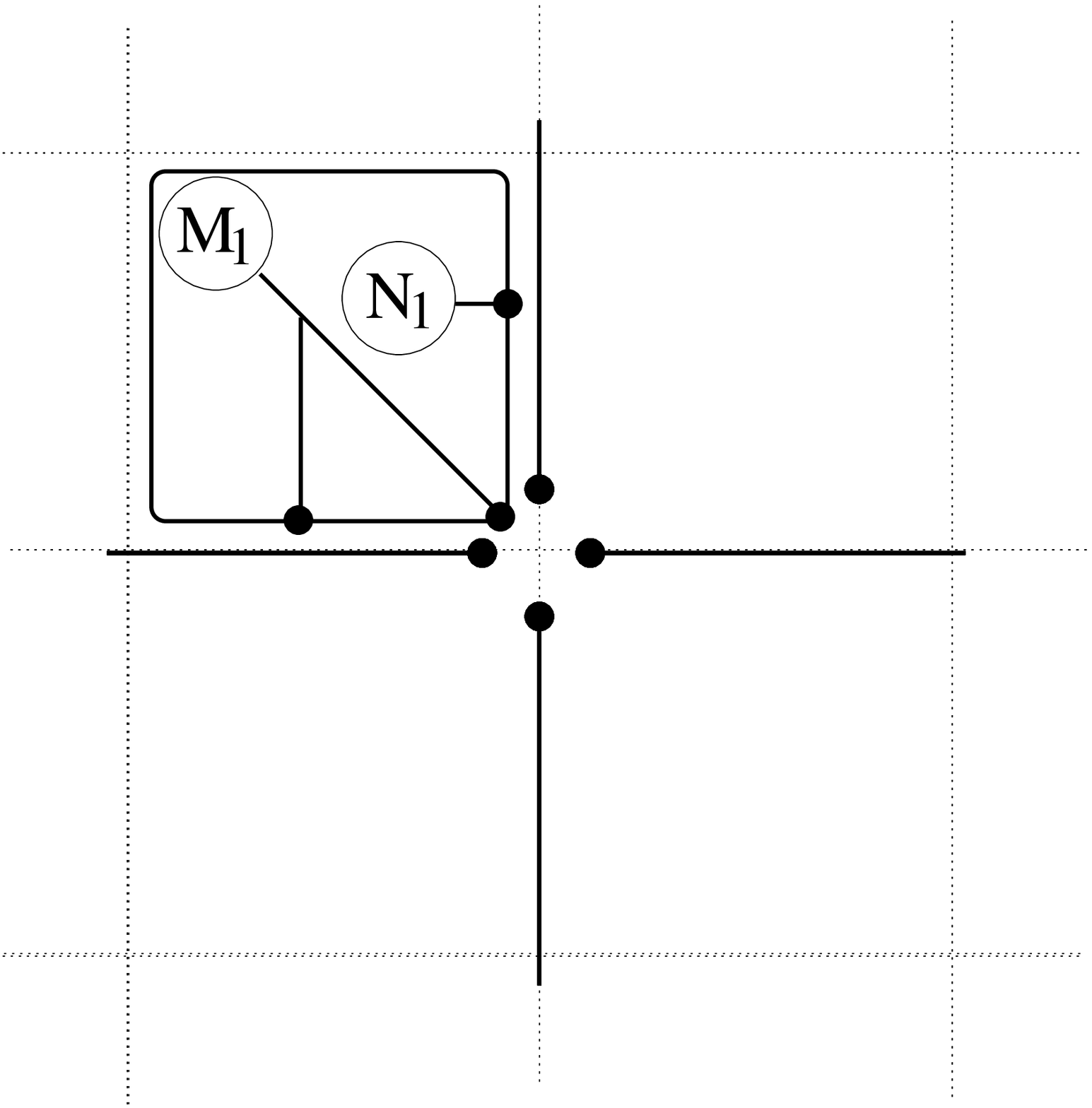}\end{array}+
\end{eqnarray*}
\begin{eqnarray*}
&&+\frac{1}{2} \begin{array}{c}  \includegraphics[width=3.4cm,angle=360]{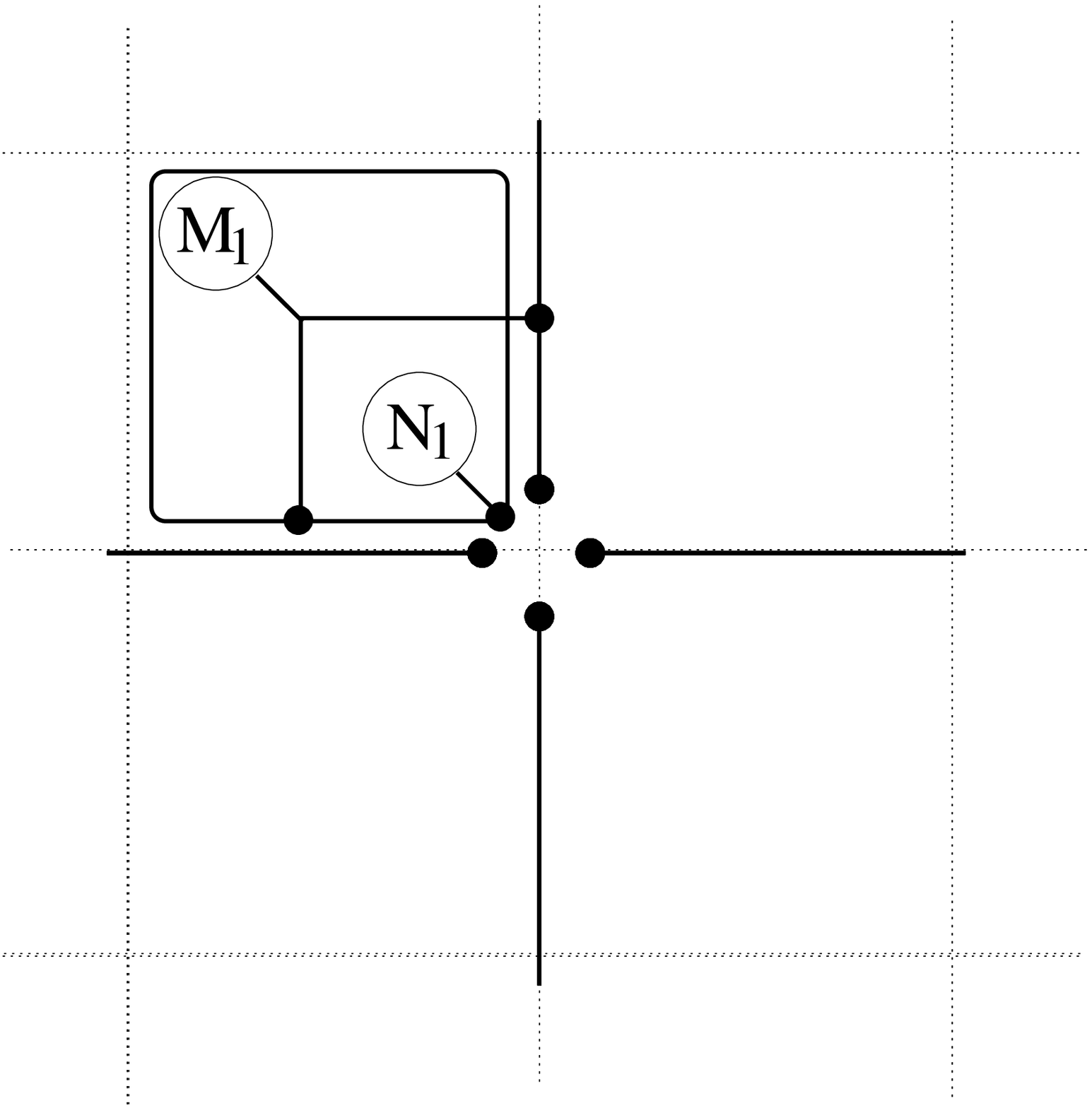}\end{array}+\frac{1}{2}\begin{array}{c}\includegraphics[width=3.4cm,angle=360]{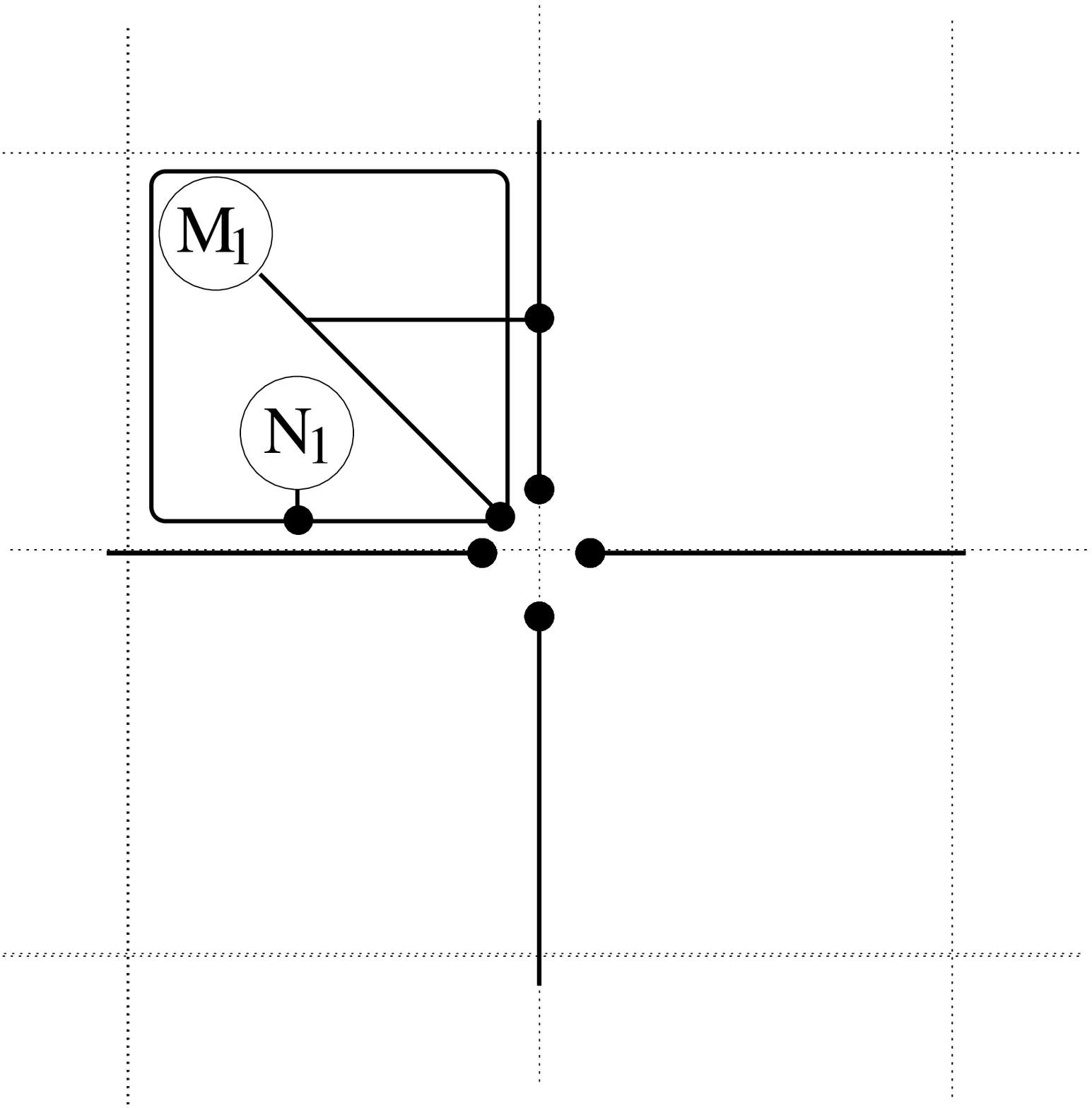}\end{array}+
\\ &&-\frac{1}{2} \begin{array}{c}  \includegraphics[width=3.4cm,angle=360]{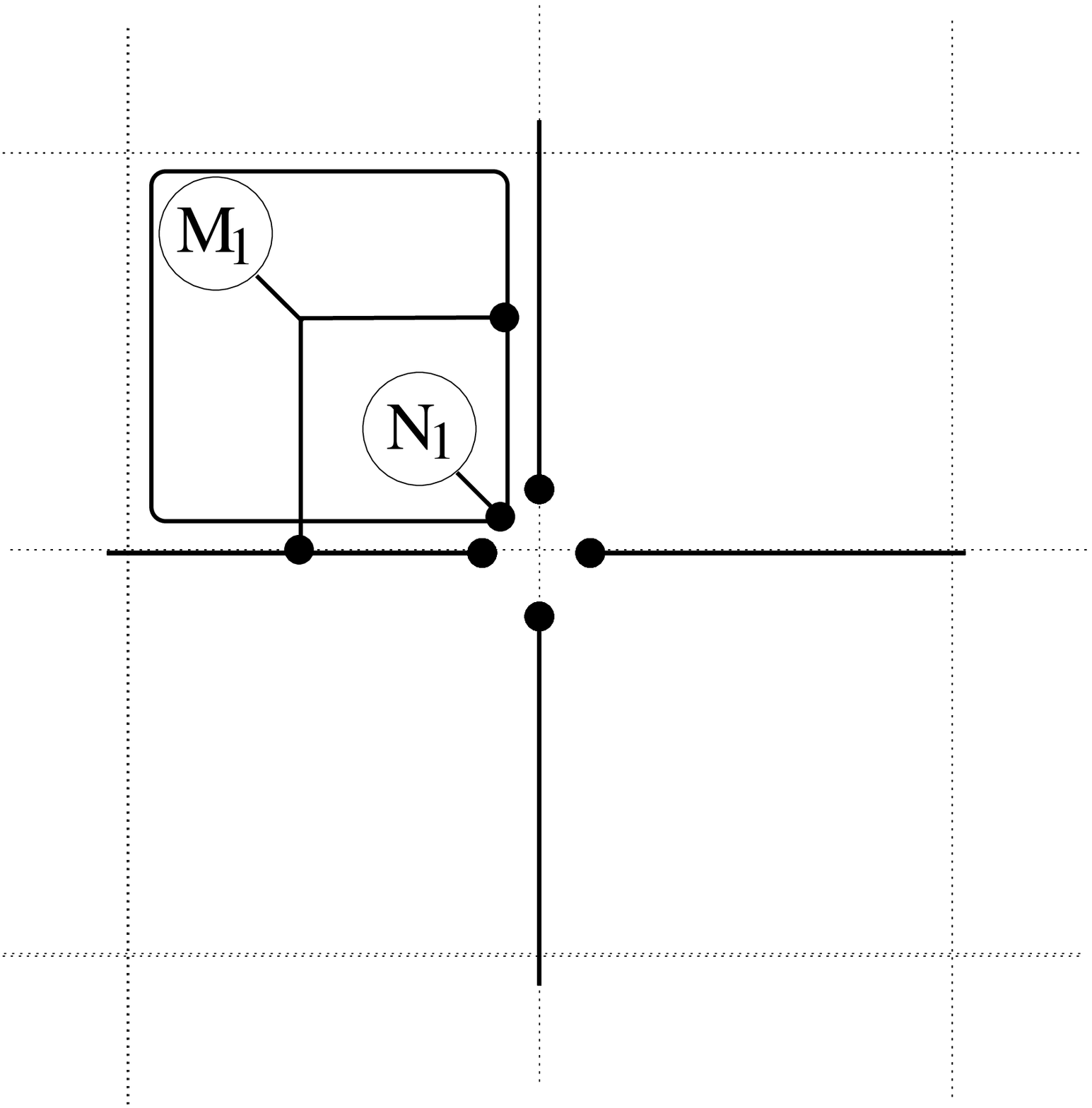}\end{array}-\frac{1}{2}\begin{array}{c}\includegraphics[width=3.4cm,angle=360]{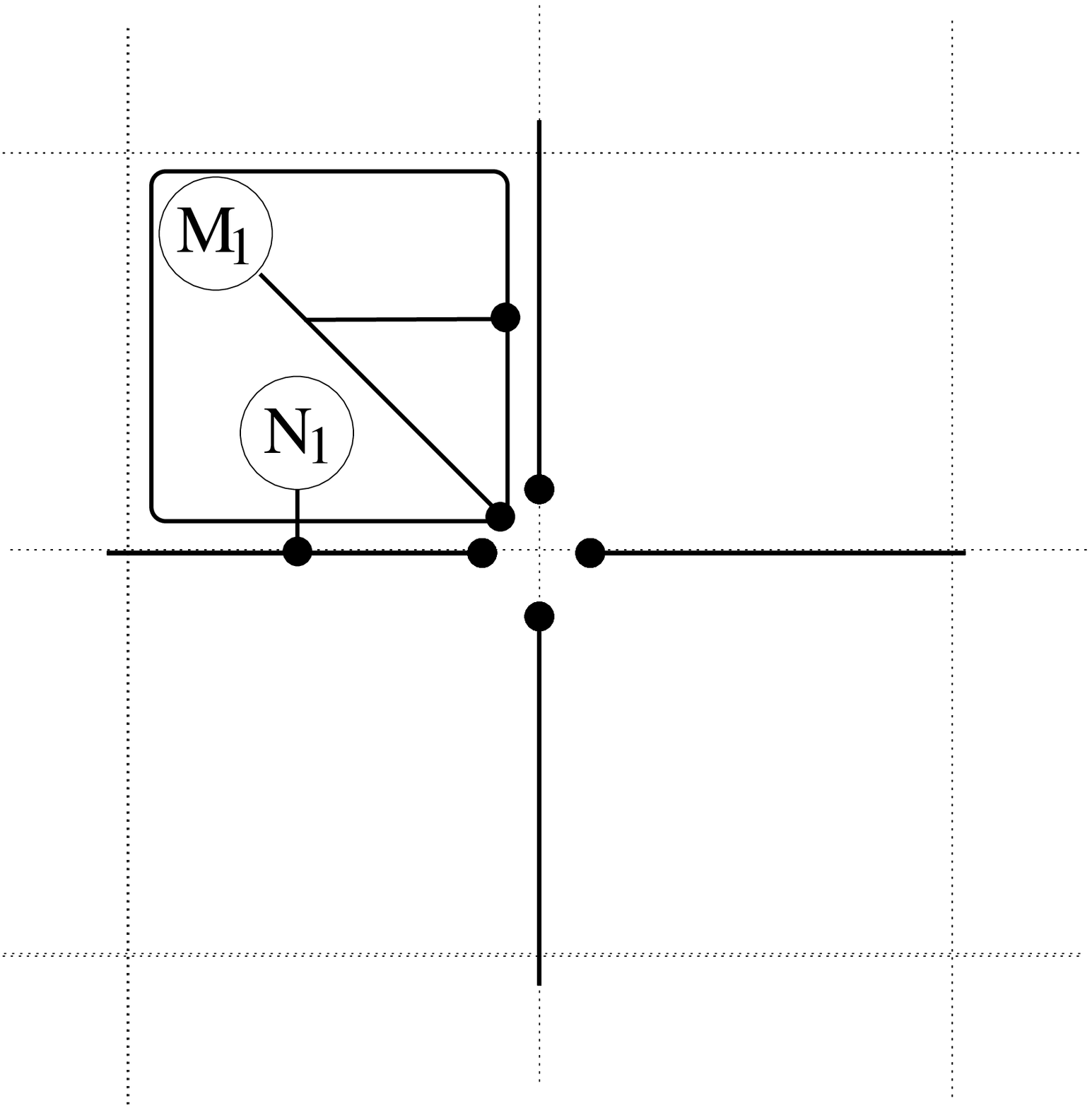}\end{array}- (\mbox{same diagrams switching $N_1\leftrightarrow M_1$})\,,
\end{eqnarray*}
where we have used the Leibnitz rule when the flux operators act on products of holonomies, and we have symmetrized the local action of the graspings at the node.
More precisely, when the product of $E$-fields appearing in $E^{\va R}(\Lambda N)$ act at a given point there is, on the one hand, a factor ordering ambiguity associated to the non-commutativity of the flux operators, and, on the other hand, an ambiguity associated to the place at which the smearing field $N$ is contracted with the Wilson line appearing in the regularization of $F^{\va R}(N)$.
By symmetrization we mean that every time we find this ambiguity we simply sum over all ordering possibilities with equal weights and divide by their number. This is the origin of the
weight factors in the previous equation.

It is easy to see that the first three diagrams in the previous equation give an amplitude that is symmetric under  $N_1\leftrightarrow M_1$ and so are canceled. This is a consequence of the symmetrized action  of graspings acting at the same point discussed above. It is important to notice that if we did not make this choice of (symmetrized) action we would immediately get contributions to the commutator that do not vanish when acting on gauge invariant states! The avoidance of this obvious anomaly justifies that above choice which is in addition very natural. Thus only the contributions of the last four diagrams remain. The result is:
\begin{eqnarray}&&
[C(N^1), C(M^1)]\rhd \Psi=\nonumber \\ && = \Lambda\, \frac{(M^1_{\alpha} N^1_{\beta}-N^1_{\alpha} M^1_{\beta})}{8}(
{\rm tr}[\{\tau^j,\tau^\beta\} W^1]\,h_1\,\epsilon_{\alpha jk}\tau^k h_2\otimes h_3\otimes h_4-
{\rm tr}[\{\tau^k,\tau^\beta\} W^1]\,\epsilon_{\alpha jk}\tau^j h_1\otimes h_2\otimes h_3\otimes h_4)\nonumber \\
&&= \frac{\Lambda}{8}\,{\rm tr}[W^1]\,(([N^1,M^1]h_1)\otimes h_2\otimes h_3\otimes h_4+h_1\otimes ([N^1,M^1]h_2)\otimes h_3\otimes h_4)\,,
\label{eq:C1C1} \end{eqnarray}
where we have used  $\{\tau^i,\tau^j\}=-\frac{1}{2}\delta^{ij}$ in the last line. An analogous calculation shows that:
\begin{equation}
[C(N^2), C(M^2)]\rhd \Psi=\frac{\Lambda}{8}\,{\rm tr}[W^2]\,(h_1\otimes ([N^2,M^2]h_2)\otimes h_3\otimes h_4+h_1\otimes h_2\otimes ([N^2,M^2]h_3)\otimes h_4)\,,
\label{eq:C2C2}
\end{equation}
\begin{equation}\label{eq:C3C3}
[C(N^3), C(M^3)]\rhd \Psi=\frac{\Lambda}{8}\,{\rm tr}[W^3]\,(h_1\otimes h_2 \otimes([N^3,M^3]h_3)\otimes h_4+h_1\otimes h_2\otimes h_3\otimes ([N^3,M^3]h_4))\,,
 \end{equation}
\begin{equation}\label{eq:C4C4}
[C(N^4), C(M^4)]\rhd \Psi=\frac{\Lambda}{8}\,{\rm tr}[W^4]\,(([N^4,M^4]h_1)\otimes h_2\otimes h_3 \otimes h_4+h_1\otimes h_2\otimes h_3 \otimes ([N^4,M^4]h_4))\,.
\end{equation}
We have now to compute the commutators among constraints in neighboring plaquettes. Let us start with $[C(N^1), C(M^2)]$:

\begin{eqnarray*}
&& [C(N^1), C(M^2)]\rhd \Psi=\left([\left[F[N^1],E[\Lambda M^2]\right]+\left[E[\Lambda N^1],F[M^2] \right]\right)\rhd \Psi=
\end{eqnarray*}
\begin{eqnarray*}
&&-\begin{array}{c}  \includegraphics[width=3.4cm,angle=360]{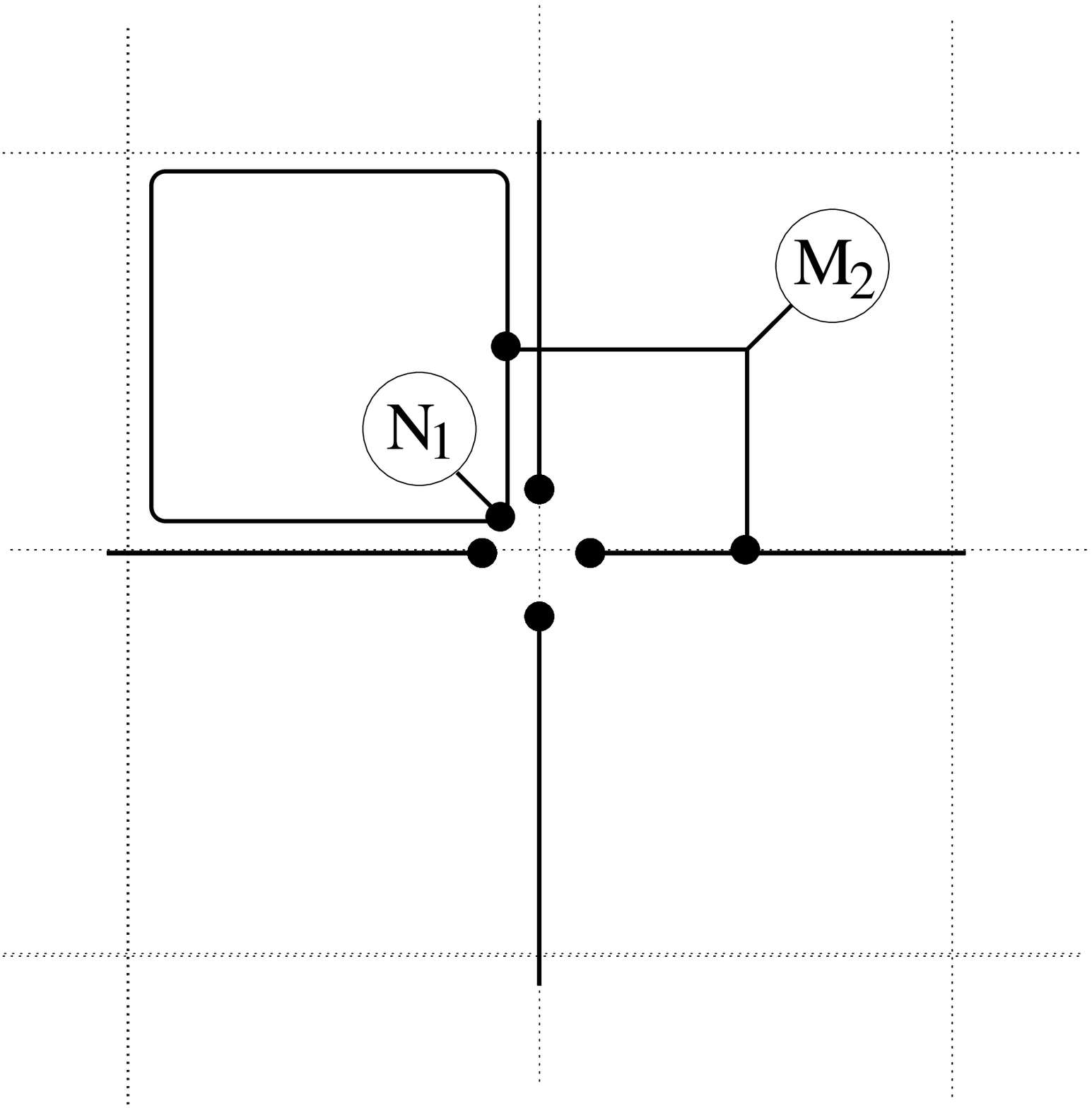}\end{array}-\begin{array}{c}  \includegraphics[width=3.4cm,angle=360]{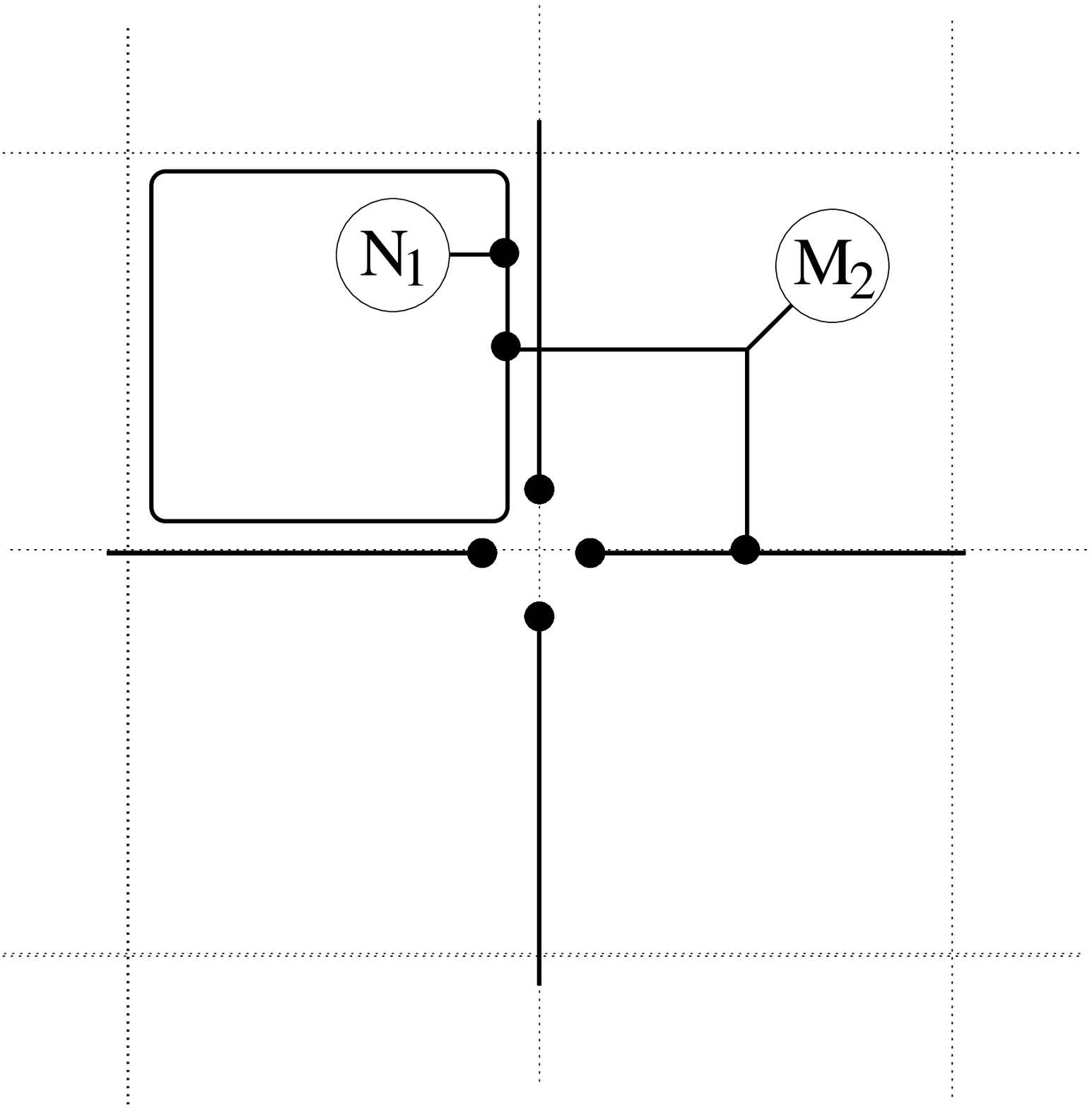}\end{array} -\begin{array}{c}\includegraphics[width=3.4cm,angle=360]{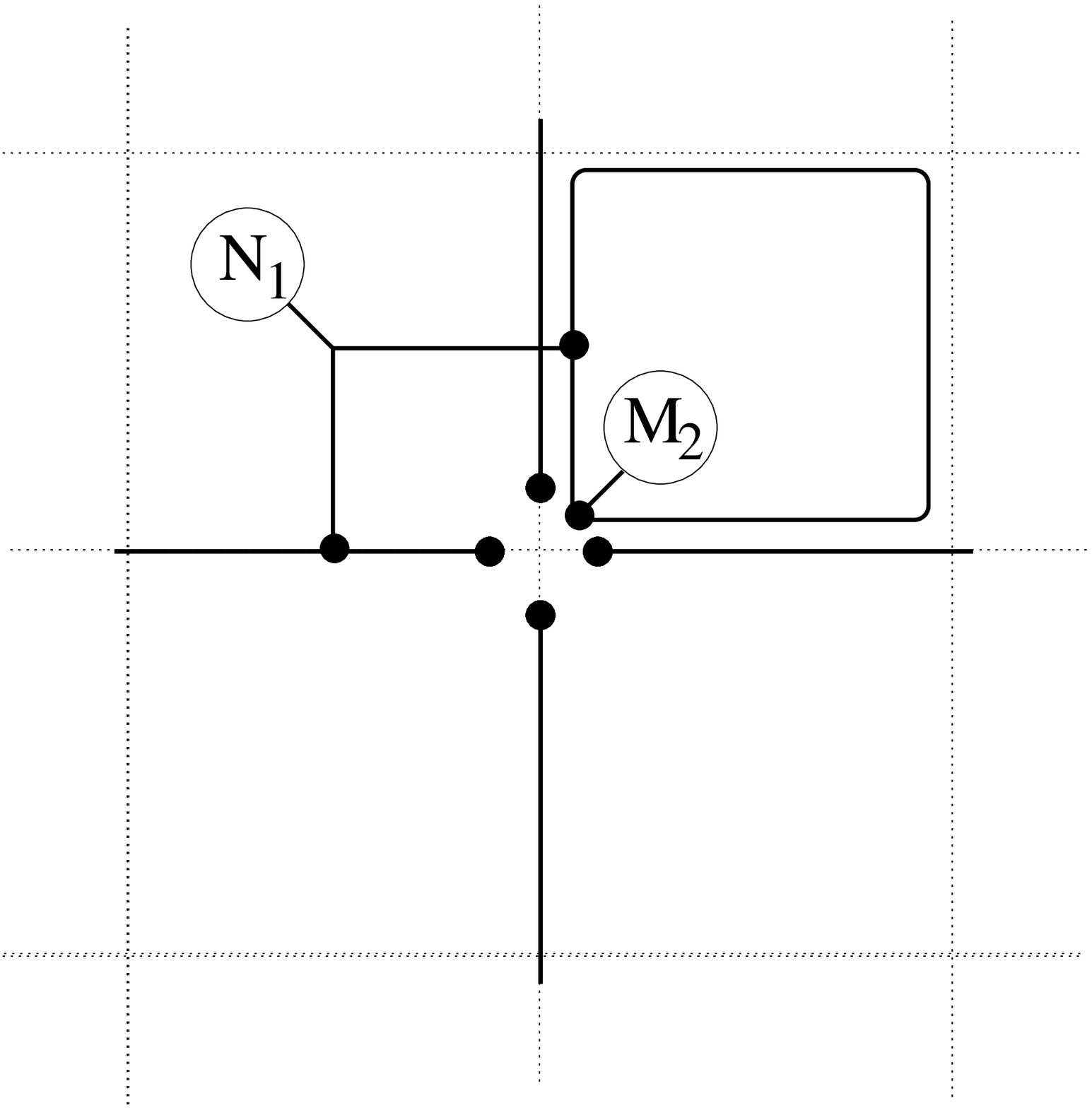}\end{array}-\begin{array}{c}  \includegraphics[width=3.4cm,angle=360]{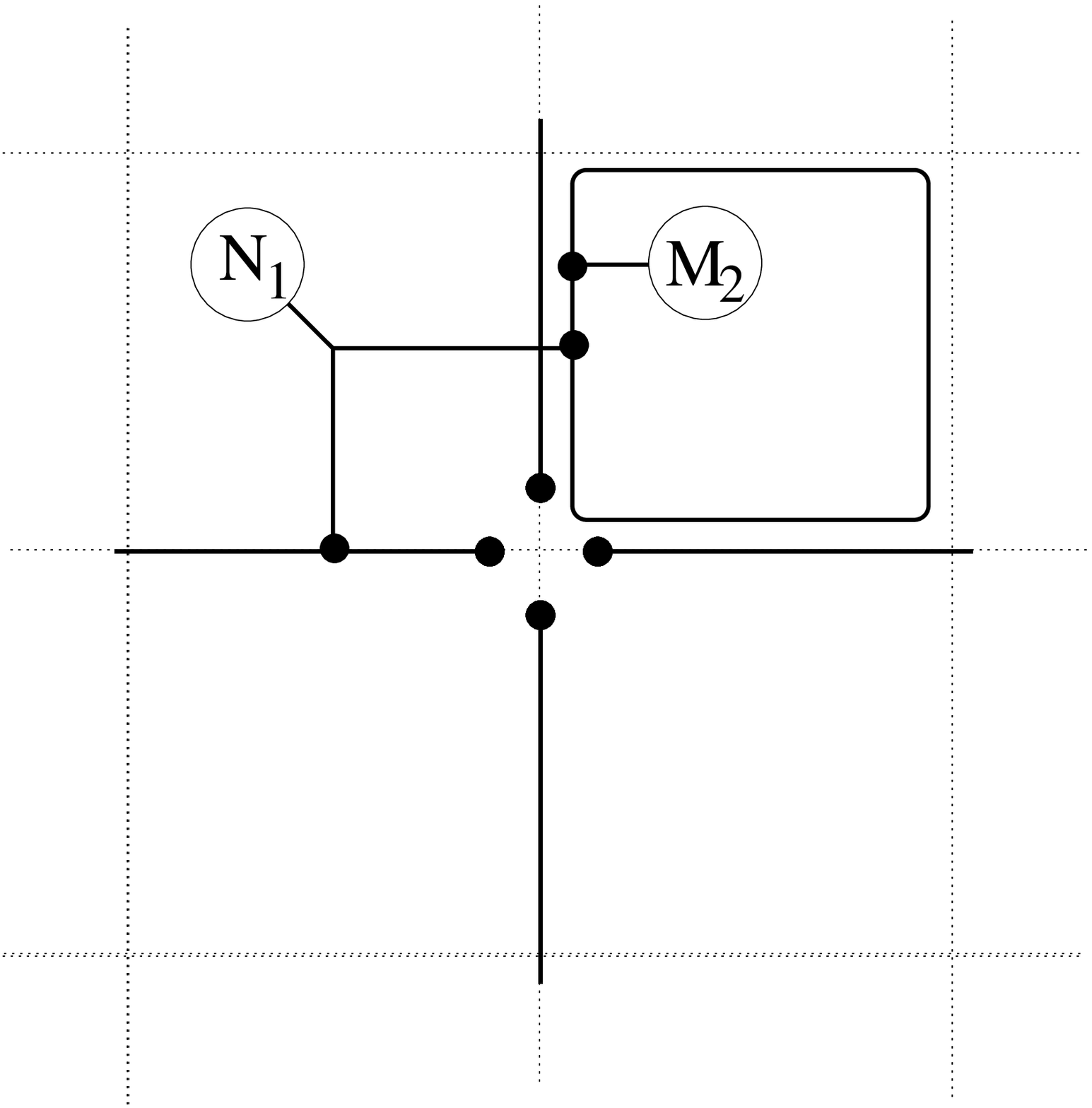}\end{array}\\ &&
=-\frac{\Lambda}{8}\, (N^1_{\alpha} M^2_\beta \ {\rm tr}[\{\tau^k,\tau^{\alpha}\} W^1]\,h_1\otimes h_2\otimes \epsilon_{\beta jk}\tau^j h_3\otimes h_4+M^2_{\alpha} N^1_\beta \ {\rm tr}[\{\tau^j,\tau^{\alpha}\} W^2]\,  \epsilon_{\beta jk}\tau^k h_1\otimes h_2\otimes h_3\otimes h_4)\\ &&=
\frac{\Lambda}{16}\, {\rm tr}[W^1]\, h_1\otimes h_2\otimes [N^1,M^2] h_3\otimes h_4+\frac{\Lambda}{16}\, {\rm tr}[W^2]\,[N^1,M^2]  h_1\otimes h_2\otimes h_3\otimes h_4).
\end{eqnarray*}
An analogous computation shows that the commutators among constraints in plaquettes which just share a vertex ($(1,3)$ and $(2,4)$) vanish, while for the other contributions we have:
\begin{equation}\nonumber
[C(N^1), C(M^4)]\rhd \Psi=\frac{\Lambda}{16}\,{\rm tr}[W^1]\,h_1\otimes h_2\otimes h_3\otimes [N^1,M^4] h_4+\frac{\Lambda}{16}\,{\rm tr}[W^4]\, h_1\otimes[N^1,M^4] h_2\otimes h_3\otimes h_4\,,
\end{equation}
\begin{equation}\nonumber
[C(N^2), C(M^1)]\rhd \Psi=\frac{\Lambda}{16}\,{\rm tr}[W^1] h_1\otimes  h_2\otimes [N^2,M^1]\, h_3\otimes  h_4+\frac{\Lambda}{16}\,{\rm tr}[W^2]\, [N^2,M^1]\, h_1\otimes h_2\otimes h_3\otimes  h_4\,,
\end{equation}
\begin{equation}\nonumber
[C(N^2), C(M^3)]\rhd \Psi=\frac{\Lambda}{16}\,{\rm tr}[W^3]\,h_1\otimes  [N^2,M^3]\,h_2\otimes  h_3\otimes  h_4+\frac{\Lambda}{16}\,{\rm tr}[W^2]\,h_1\otimes h_2\otimes h_3\otimes [N^2,M^3]\,h_4\,,
\end{equation}
\begin{equation}\nonumber
[C(N^3), C(M^2)]\rhd \Psi=\frac{\Lambda}{16}\,{\rm tr}[W^3]\,h_1\otimes  [N^3,M^2]\,h_2\otimes  h_3\otimes  h_4+\frac{\Lambda}{16}\,{\rm tr}[W^2]\,h_1\otimes h_2\otimes h_3\otimes [N^3,M^2]\,h_4\,,
\end{equation}
\begin{equation}\nonumber
[C(N^3), C(M^4)]\rhd \Psi=\frac{\Lambda}{16}\,{\rm tr}[W^3]\,[N^3,M^4]\,h_1\otimes  h_2\otimes  h_3\otimes  h_4+\frac{\Lambda}{16}\,{\rm tr}[W^4]\,h_1\otimes h_2\otimes [N^3,M^4]\,h_3\otimes  h_4\,,
\end{equation}
\begin{equation}\nonumber
[C(N^4), C(M^3)]\rhd \Psi=\frac{\Lambda}{16}\,{\rm tr}[W^3]\,[N^4,M^3]\,h_1\otimes  h_2\otimes  h_3\otimes  h_4+\frac{\Lambda}{16}\,{\rm tr}[W^4]\,h_1\otimes h_2\otimes [N^4,M^3]\,h_3\otimes  h_4\,,
\end{equation}
\begin{equation}\nonumber
[C(N^4), C(M^1)]\rhd \Psi=\frac{\Lambda}{16}\,{\rm tr}[W^1]\,h_1\otimes  h_2\otimes  h_3\otimes [N^4,M^1]\, h_4+\frac{\Lambda}{16}\,{\rm tr}[W^4]\, h_1\otimes [N^4,M^1]\,h_2\otimes h_3\otimes  h_4\,.
\end{equation}
The fact that four different values of the smearing field enter the action of the regularized constraint at a single point is an artifact of the discretization since the values of the smearing field $N$, being discretized, changes from plaquette to plaquete in the four plaquettes surrounding the node of interest. The `multi-valuedness' of smearing fields created by the discretization makes more obscure the meaning of the previous equations and should not affect the final result. A more transparent interpretation of the equations is obtained by {\em coarse graining} the discrete smearing fields (the equalities that are produced by this means will be denoted by the symbol $\cg$, where $c.g.$ stands here for coarse graining).
Therefore we can evaluate the action (\ref{eq:dinamic_discrete_left}) by the appropriate coarse-graining procedure $N^1=N^2=N^3=N^4\equiv N$, $M^1=M^2=M^3=M^4\equiv M$ and see that at the quantum level the algebra of the curvature constraint with itself presents an anomaly, namely:
\begin{eqnarray}\label{eq:CCanomaly}
&&[C^{\va R}[N],C^{\va R}[M]]\rhd\Psi\cg \nonumber \\ && \cg \frac{\Lambda}{8} ({\rm tr}[W^1]+{\rm tr}[W^2]+{\rm tr}[W^3]+{\rm tr}[W^4])
\left[([N,M]h_1)\otimes h_2\otimes h_3\otimes h_4\right. +\nonumber \\ && \nonumber + \left.h_1\otimes ([N,M]h_2)\otimes h_3\otimes h_4+
h_1\otimes h_2\otimes ([N,M]h_3)\otimes h_4+h_1\otimes h_2\otimes h_3\otimes ([N,M]h_4)\right]=\nonumber\\
&&={\Lambda} \,G^{\va R}(\frac{{\rm tr}[W]}{2}[N,M])\rhd\Psi~,
\end{eqnarray}
where $G^{\va R}(\frac{{\rm tr}[W]}{2}[N,M])=\sum_v(\sum_{p\in v}{\rm tr}[W_p]/8)(\sum_{p\in v}G^p[N,M])$.
Notice that the classical analog reduces---once the regulator is removed ($\epsilon\rightarrow 0$) for smooth field configurations and due to the fact that ${\rm Tr}[W]/2\to 1$---to the correct algebra: \[ \{C[N],C[M]\}=\Lambda \,G(\{N,M\}).\] The presence of ${\rm tr}[W]$ in the regulated algebra will not disappear in the
quantum case: this is a genuine quantization anomaly introduced by our regularization method.

We can now compute the action on the state $\Psi$ of the commutator between the scalar and the Gauss constraints and verify that the relation
\begin{equation}\label{eq:Gauss_discrete}
[C^{\va R}(N),G^{\va R}(M)]=C^{\va R}([N,M])
\end{equation}
	holds. In this case when acting with the lhs of (\ref{eq:Gauss_discrete}) on a single node, for the curvature constraint there will be four relevant terms in the sum over plaquettes entering the definition of $C^{\va R}(N)$ (the terms related to the four plaquettes around the given node), each with a different value of the smearing field $N^p$, while for the Gauss constraint there is only one relevant term (the one related to the dual plaquette around the given node) and we will denote $M^{p*}$ the value of the smearing field associated to this term. Thereby, for the lhs of (\ref{eq:Gauss_discrete}) there are four commutators to compute and for each of them there are two terms: one related to the $F^{\va R}$ part of $C^{\va R}$ and one to the $E^{\va R}$ part.
	
Since the $F(N^p)$, for all the four plaquettes ($p=1,2,3,4$) around the given node, depend only on holonomies the Leibnitz rule applied to the their commutators with the Gauss constraint in the dual plaquette around the given node $G[M^{p*}]$ implies directly
\be\label{fifi}
\left[F^{\va R}(N),G^{\va R}(M)\right]\rhd\Psi=F^{\va R}([N,M])\rhd\Psi\,.
\ee
This implies that we can simply concentrate on the commutator of $E^{\va R}(\Lambda N)$ with the Gauss constraint.

Starting with the plaquette $1$, we have, omitting the index $p*$ for the smearing field $M^{p*}$:
\ba
&& \nonumber  [E(\Lambda N^1),G(M)]\rhd\Psi=\\
&& = \Lambda \begin{array}{c}  \includegraphics[width=4cm,angle=360]{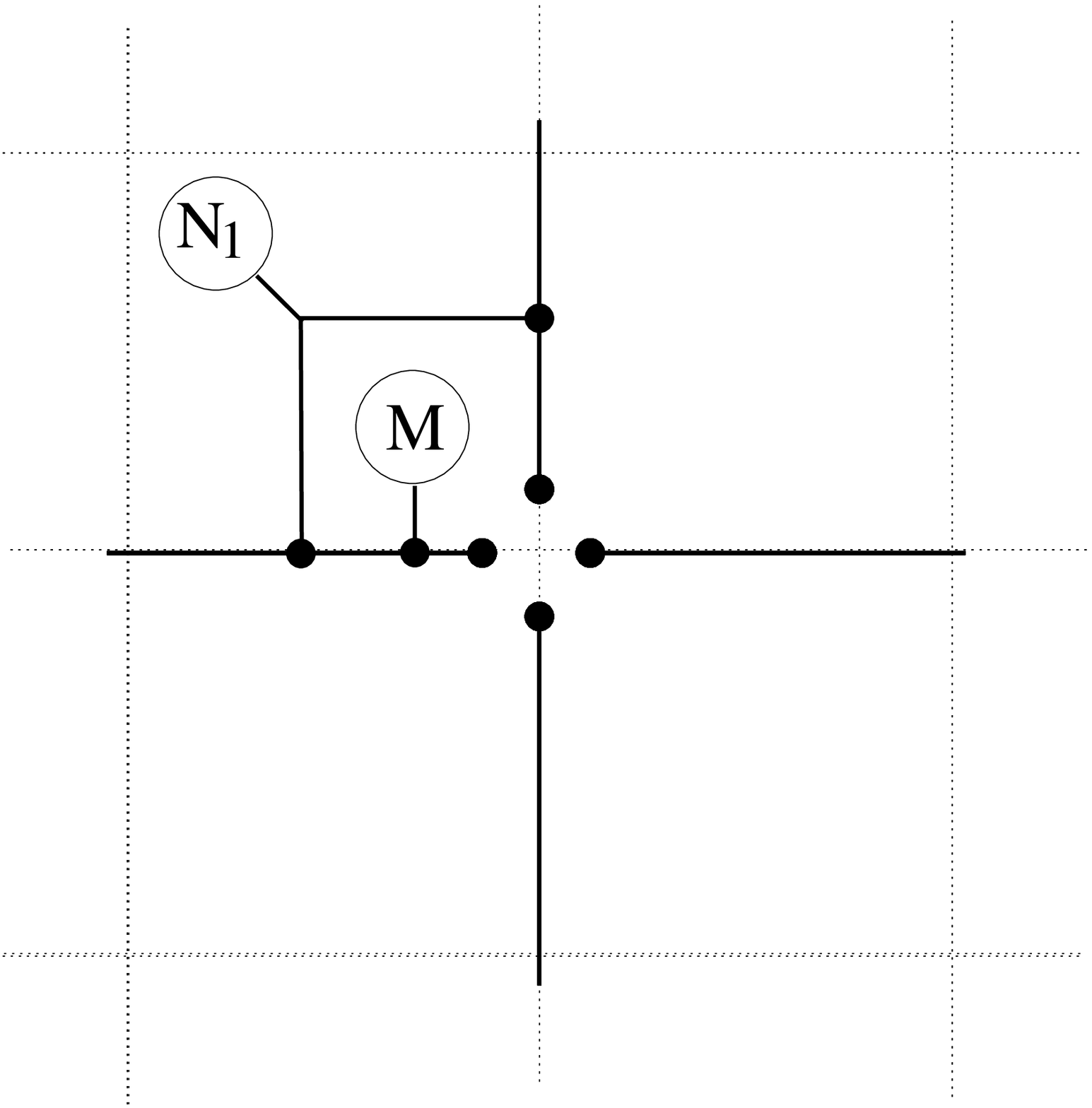}\end{array}-\Lambda \begin{array}{c}  \includegraphics[width=4cm,angle=360]{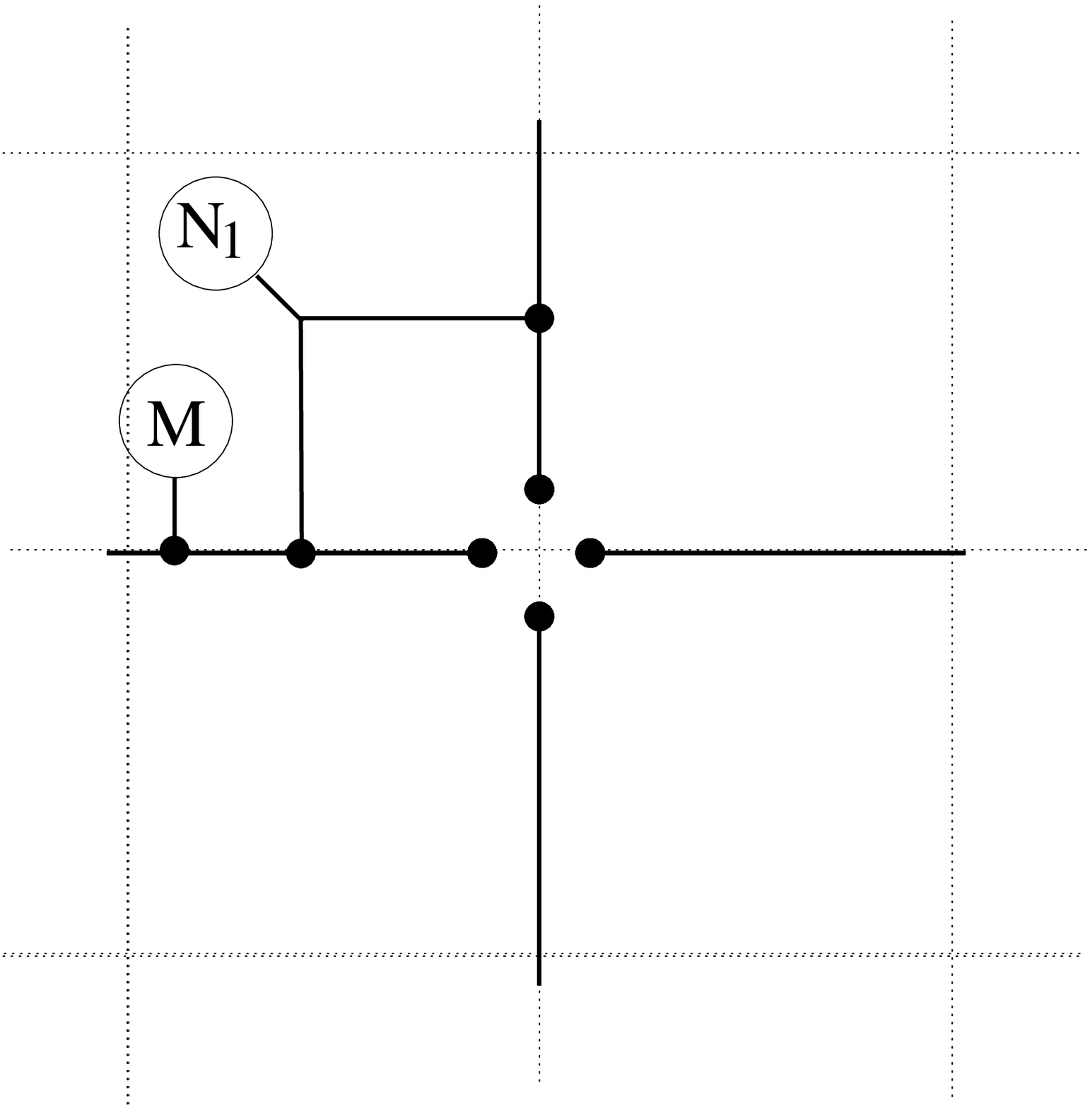}\end{array}+\nonumber \\ &&
+\Lambda \begin{array}{c}  \includegraphics[width=4cm,angle=360]{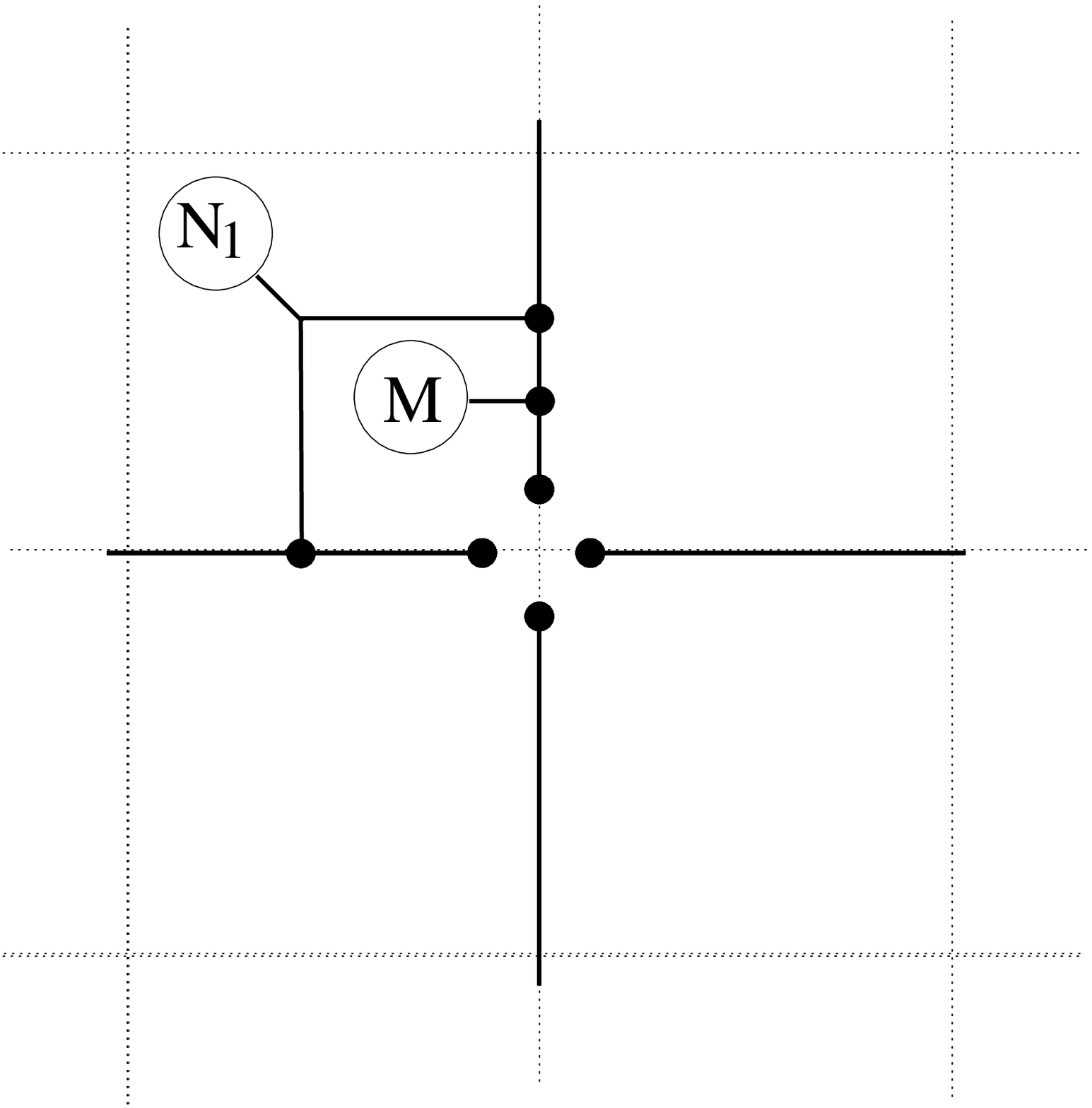}\end{array}-\Lambda \begin{array}{c}  \includegraphics[width=4cm,angle=360]{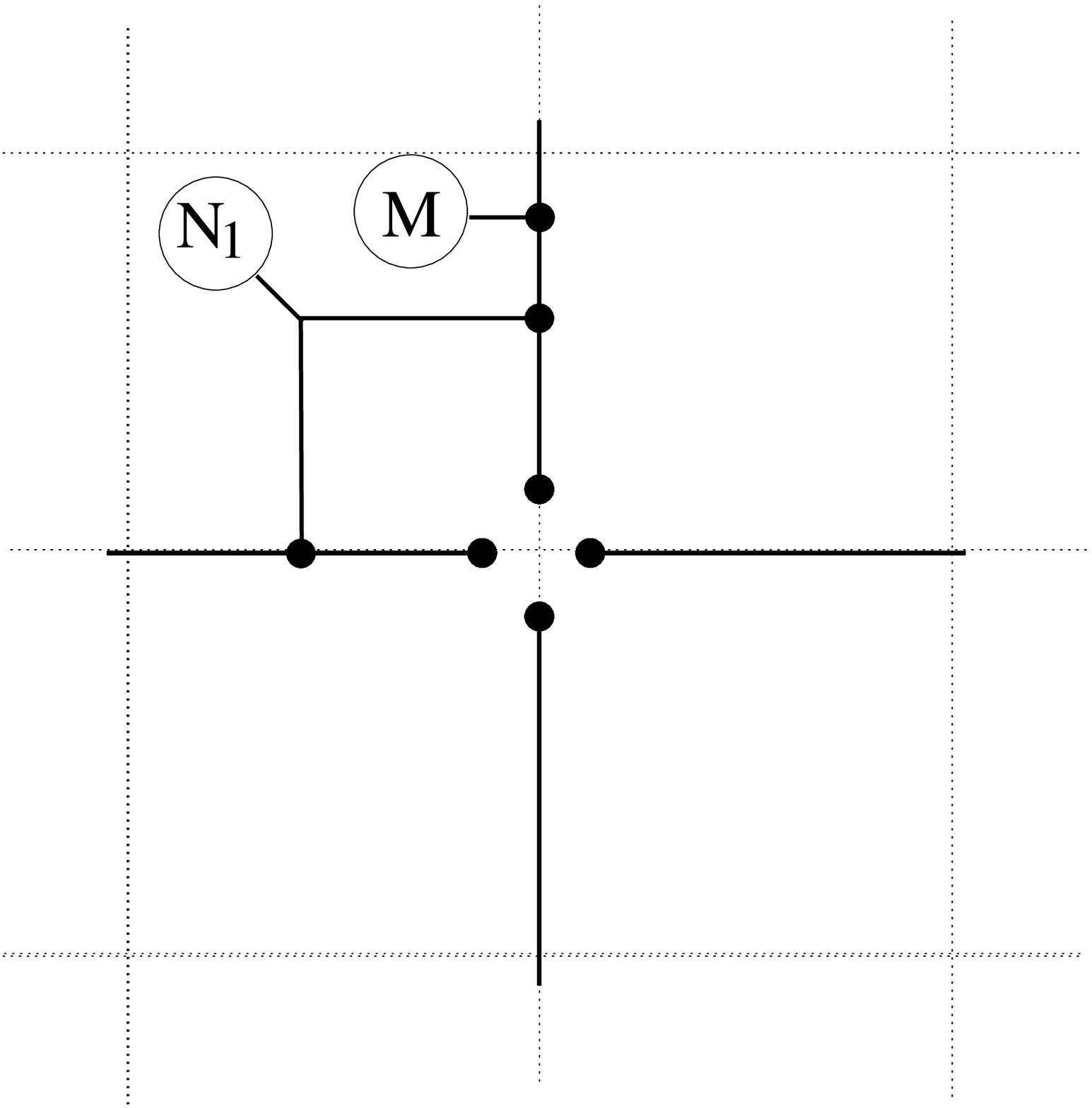}\end{array}
\ea
Simple algebra yields\begin{equation}\label{eq:C1G1}
[E(\Lambda N^1),G(M)]\rhd\Psi=\frac{\Lambda}{4} N^1h_1\otimes Mh_2\otimes h_3\otimes h_4-\frac{\Lambda}{4} Mh_1\otimes N^1h_2\otimes h_3\otimes h_4,
\end{equation}
In an analogous way one can compute:
\begin{equation}\label{eq:C2G2}
[E(\Lambda N^2),G(M)]\rhd\Psi=\frac{\Lambda}{4} h_1\otimes N^2h_2\otimes Mh_3\otimes h_4-\frac{\Lambda}{4} h_1\otimes Mh_2\otimes N^2h_3\otimes h_4\,,
\end{equation}
\begin{equation}\label{eq:C3G3}
[E(\Lambda N^3),G(M)]\rhd\Psi=\frac{\Lambda}{4} h_1\otimes h_2\otimes N^3h_3\otimes Mh_4-\frac{\Lambda}{4} h_1\otimes h_2\otimes Mh_3\otimes N^3h_4\,,
\end{equation}
\begin{equation}\label{eq:C4G4}
[E(\Lambda N^4),G(M)]\rhd\Psi=\frac{\Lambda}{4} Mh_1\otimes h_2\otimes h_3\otimes N^4h_4-\frac{\Lambda}{4} N^4h_1\otimes h_2\otimes h_3\otimes Mh_4\,.
\end{equation}
We can now sum up all the contribution and we get:
\begin{eqnarray}\label{eq:Gauss_discrete_left}
&& [E^{\va R}(\Lambda N),G^{\va R}(M)]\rhd\Psi=\nonumber\\
&&+\frac{\Lambda}{4} N^1h_1\otimes Mh_2\otimes h_3\otimes h_4-\frac{\Lambda}{4} Mh_1\otimes N^1h_2\otimes h_3\otimes h_4+\nonumber\\
&&+\frac{\Lambda}{4} h_1\otimes N^2h_2\otimes Mh_3\otimes h_4-\frac{\Lambda}{4} h_1\otimes Mh_2\otimes N^2h_3\otimes h_4+\nonumber\\
&&+\frac{\Lambda}{4} h_1\otimes h_2\otimes N^3h_3\otimes Mh_4-\frac{\Lambda}{4} h_1\otimes h_2\otimes Mh_3\otimes N^3h_4+\nonumber\\
&&+ \frac{\Lambda}{4} Mh_1\otimes h_2\otimes h_3\otimes N^4h_4-\frac{\Lambda}{4} N^4h_1\otimes h_2\otimes h_3\otimes Mh_4~.
\end{eqnarray}
We have now to compute the r.h.s of (\ref{eq:Gauss_discrete}). Starting again from the plaquette $1$,we get:
\begin{eqnarray}
\label{eq:CG1}
&& E(\Lambda [N^1,M])\rhd\Psi=
\frac{\Lambda}{4} N^1_i M_j \epsilon_{ijk}\,\epsilon^k_{\ rs}\ \tau^r h_1\otimes \tau^s h_2\otimes h_3\otimes h_4=\nonumber\\
&&= \frac{\Lambda}{4}  N^1 h_1\otimes M h_2\otimes h_3\otimes h_4-\frac{\Lambda}{4}  M h_1\otimes N^1 h_2\otimes h_3\otimes h_4\,.
\end{eqnarray}
An analogous computation shows that:
\begin{eqnarray}\label{eq:CG2}
E(\Lambda [N^2,M])\rhd\Psi=
 \frac{\Lambda}{4}  h_1\otimes N^2 h_2\otimes M h_3\otimes h_4-\frac{\Lambda}{4}   h_1\otimes M h_2\otimes N^2 h_3\otimes h_4\,,
\end{eqnarray}
\begin{eqnarray}\label{eq:CG3}
E(\Lambda [N^3,M])\rhd\Psi=
 \frac{\Lambda}{4}  h_1\otimes h_2\otimes N^3 h_3\otimes M h_4-\frac{\Lambda}{4}  h_1\otimes h_2\otimes Mh_3\otimes N^3h_4\,,
\end{eqnarray}
\begin{eqnarray}\label{eq:CG4}
E(\Lambda [N^4,M])\rhd\Psi=
 \frac{\Lambda}{4}  M h_1\otimes  h_2\otimes h_3\otimes N^4 h_4-\frac{\Lambda}{4}  N^4 h_1\otimes  h_2\otimes h_3\otimes Mh_4\,.
\end{eqnarray}
Thus, summing all the four contributions to the action of the r.h.s of (\ref{eq:Gauss_discrete}) on $\Psi$, we get:
\begin{eqnarray}\label{eq:Gauss_discrete_right}
&& E^{\va R}(\Lambda [N,M])\rhd\Psi= \nonumber\\
&&+\frac{\Lambda}{4} N^1h_1\otimes Mh_2\otimes h_3\otimes h_4-\frac{\Lambda}{4} Mh_1\otimes N^1h_2\otimes h_3\otimes h_4+\nonumber\\
&&+\frac{\Lambda}{4} h_1\otimes N^2h_2\otimes Mh_3\otimes h_4-\frac{\Lambda}{4} h_1\otimes Mh_2\otimes N^2h_3\otimes h_4+\nonumber\\
&&+\frac{\Lambda}{4} h_1\otimes h_2\otimes N^3h_3\otimes Mh_4-\frac{\Lambda}{4} h_1\otimes h_2\otimes Mh_3\otimes N^3h_4+\nonumber\\
&&+ \frac{\Lambda}{4} Mh_1\otimes h_2\otimes h_3\otimes N^4h_4-\frac{\Lambda}{4} N^4h_1\otimes h_2\otimes h_3\otimes Mh_4~.
\end{eqnarray}

From eq.(\ref{eq:Gauss_discrete_left}) and (\ref{eq:Gauss_discrete_right}) we conclude that $[E^{\va R} (\Lambda N),G^{\va R}(M)]= E^{\va R}(\Lambda [N,M])$ which combined with (\ref{fifi}) yields
\be
[C^{\va R}(N),G^{\va R}(M)]= C^{\va R}([N,M]),
\ee
as expected.

To summarize, we have seen that the quantum version of the constraints algebra (\ref{eq:constraints_algebra}) of gravity in $2+1$ dimensions with non-vanishing cosmological constant reads:
\begin{eqnarray*}
[C^{\va R}(N),C^{\va R}(M)]\cg {\Lambda} \,G^{\va R}(\frac{{\rm tr}[W]}{2}[N,M])
\end{eqnarray*}
\begin{eqnarray*}
[G^{\va R}(N),G^{\va R}(M)]= G^{\va R}([N,M])
\end{eqnarray*}
\begin{eqnarray}\label{eq:quantum constraints_algebra}
[C^{\va R}(N),G^{\va R}(M)]=  C^{\va R}([N,M])~
\end{eqnarray}
Relations (\ref{eq:quantum constraints_algebra}) show that just the commutators among the scalar constraints present an anomaly due to the presence of the factor $\frac{{\rm tr}[W]}{2}$ in the smearing of the
Gauss law on the r.h.s. of the first equation. We see that the regularization does not break the internal gauge group $SU(2)$; however, it does break the part of the gauge symmetry group related to spacetime
 diffeomorphisms.  Notice also that the anomaly is a genuine quantum effect. If we had computed the Poisson algebra of regularized constraints instead we would have found basically the same result (where commutators are replaced by Poisson brackets). However, in that case the problematic factor disappears in the continuum limit as $\frac{{\rm tr}[W]}{2}=1+\sO(\epsilon^4)$.

It is useful to rewrite the result in terms of a constraint reparametrization that exhibits more clearly the compact character of the gauge symmetry group of Riemannian
2+1 gravity with positive cosmological constant.  At the classical level we know that the algebra (\ref{eq:constraints_algebra}) generates a local $\emph{su(2)}\otimes\emph{su(2)}$ symmetry, as one can see immediately by defining $F^{\pm}(N)=C(N)\pm\sqrt{\Lambda}\,G(N)$ and computing:
\begin{equation}\label{eq:classical symmetry}
\{F^{\pm}(N),F^{\pm}(M)\}= \pm2\sqrt{\Lambda}\,F^{\pm}([N,M]) ~~~~~~~~~~\{F^{\pm}(N),F^{\mp}(M)\}=0~.
\end{equation}
In order to see what modification to this local symmetry appears at the quantum level we define the operators $F^{{\va R} \pm}(N)=C^{\va R}(N)\pm\sqrt{\Lambda}\,G^{\va R}(N)$ and then compute the commutators $[F^{{\va R}\pm}(N),F^{{\va R} \pm}(M)]$ and $[F^{{\va R} \pm}(N),F^{{\va R}\mp}(M)]$. The result is
\begin{eqnarray}\label{eq:quatum symmetry1}
[F^{{\va R} \pm}(N),F^{{\va R} \pm}(M)]=\pm2\sqrt{\Lambda}\,F^{{\va R} \pm}([N,M])
+\frac{\Lambda}{2} G^{\va R}(({{\rm tr} [W]}-2)[N,M])
\end{eqnarray}
and
\begin{eqnarray}\label{eq:quatum symmetry2}
[F^{{\va R} \pm}(N),F^{{\va R}\mp}(M)]&=&\frac{\Lambda}{2}\,G^{\va R}(( {{\rm tr} [W]}-2)[N,M]).
\end{eqnarray}

 \section{Discussion}

We have precisely computed the regulated quantum constraint algebra for Riemannian $2+1$ gravity in the connection formulation of LQG.
 The nature of the kinematical Hilbert space of LQG imposes the need of a regularization in the definition of the quantum constraints.
 This is so, due to the fact that only the holonomy (and not the local connection) and conjugate fluxes (instead of the local triad) can be quantized in the kinematical LQG representation: the fundamental operators representing phase space variables are of extended nature.  We have studied a simple regularization of the constraints that leads to the
 correct naive continuum limit of both the constraints and their Poisson algebra. However, when these regulated constraints are quantized
the regulated quantum constraint algebra becomes anomalous due to the presence of plaquette loop operators that do not go away in the refinement limit
of the regulating lattice \footnote{We would like to point to the existence of results along the lines of this work but in 3+1 dimensions by R. Loll \cite{loll} where
the 3d diffeomorphism constraint sub-algebra is tested in a lattice regularization of quantum gravity. }.

 There is a large freedom in the choice of regularization of the constraints. One source of ambiguity comes from the fact that the flux operators corresponding to the
 local triad do not commute in the quantum theory. As the classical constraints of the theory considered here are non linear in the triad field---due to the presence of a non vanishing cosmological constant---this non-commutativity introduces factor ordering ambiguities. We have shown that the requirement that the quantum constraint algebra be satisfied completely eliminates this source of ambiguity. More precisely, the caculations leading to eq. (\ref{eq:C1C1}) and finally to eq. (\ref{eq:CCanomaly}) show that if we do not symmetrize over all possible orderings then $[C^{\va R}(N), C^{\va R}(M)]\rhd \Psi\not=0$ when acting on states $\Psi$ annihilated by the quantum Gauss constraints.

Another source of ambiguity can be found in the choice of paths along which the fluxes and holonomies used in the regularization are defined. For simplicity we have chosen here
a square lattice $\CS$ and its dual. However, it  should be clear from our calculation that the presence of the anomaly found here is independent of this choice. It cannot be removed by playing with this freedom. Finally, there is the ambiguity in the choice of representation used in the regularization of the curvature constraint. Instead of regularizing the local curvature using the fundamental representation as in (\ref{curvy}) we could have used the more general function
\be f[NW]=\left.\frac{\sum_{j}^{n/2} {c_j} \frac{\chi_j(NW)}{j(j+1) (2j+1)}}{\sum_{k}^{n/2} c_k}\right|_{W=1+\epsilon^2 F}=\epsilon^2 N_iF^i+\sO (\epsilon^2),\ee
 for arbitrary coefficients $c_j$ and some positive integer $n$. Calculations along the lines presented here show that
 it is not possible to correct the anomaly found here by playing with the parameters of this space of functions.

 Let us mention that the anomaly found here seem to be related to the results obtain in reference \cite{Bianca} by a different approach.
 If we write the constraint algebra as in (\ref{eq:quatum symmetry1}) and (\ref{eq:quatum symmetry2}) the anomaly is parametrized by the local quantity
 $({\rm Tr}[W]-2)$. If for a moment we think of this factor as a classical quantity, and we evaluate it on a smooth connection field configuartion, then this factor is proportional to
 the local curvature squared of the connection. This is in direct correspondence with the results of \cite{Bianca}.

The anomaly found here represents an important obstacle for the implementation of the standard Dirac quantization in the LQG representation.
In particular the group averaging techniques to solve the first class constraints at the quantum level, used in the $\Lambda=0$ case, are not available.
It may be possible to contour this obstacle by means of some alternative formulations developed for the quantization of systems
with constraint algebras which are not associated with a structure Lie group (e.g. Thiemann et al. master constraint program \cite{Thiemann:2003zv}).
At this point we can only speculate with the possibility that the anomaly found here may be at the end related with the deformation of the classical symmetry
of gravity leading to the quantum group structure underlying the quantization of 2+1 gravity with non-vanishing cosmological constant found by other methods.

\section*{Acknowledgements}

We would like to thank  Bianca Dittrich,  Antonino Marcian\'o, and Michael Reisenberger  for many very useful discussions
on issues related to the subject of this paper and Eugenio Bianchi, Carlo rovelli, and Simone Speziale for suggesting a way to improve the regularization of the Gauss law. This work was supported in part by the
Agence Nationale de la Recherche; grant ANR-06-BLAN-0050 and from  {\em
l'Institut Universitaire de France}.

\end{document}